\documentclass[twocolumn]{aastex63}
\usepackage{newtxtext,newtxmath}
\usepackage[T1]{fontenc}
\usepackage[T1]{fontenc}
\usepackage{CJKutf8}
\usepackage{ae,aecompl}
\usepackage{graphicx}
\usepackage{amsmath}
\usepackage{amssymb}	
\usepackage{float}
\usepackage[]{hyperref}
\newcommand{\ud}{\mathrm{d}}

\usepackage{booktabs}
\received{November 2, 2022}
\revised{March 14, 2023}
\accepted{March 28, 2023}

\begin{document}

\title{Unraveling Twisty Linear Polarization Morphologies in Black Hole Images}


\shorttitle{Linear Polarization}
\shortauthors{R. Emami et. al.}

\correspondingauthor{Razieh Emami}
\email{razieh.emami$_{-}$meibody@cfa.harvard.edu}

\author[0000-0002-2791-5011]{Razieh Emami}
\affiliation{Center for Astrophysics $\vert$ Harvard \& Smithsonian, 60 Garden Street, Cambridge, MA 02138, USA}

\author[0000-0001-5287-0452]{Angelo Ricarte}
\affiliation{Center for Astrophysics $\vert$ Harvard \& Smithsonian, 60 Garden Street, Cambridge, MA 02138, USA}
\affiliation{Black Hole Initiative at Harvard University, 20 Garden Street, Cambridge, MA 02138, USA}

\author[0000-0001-6952-2147]{George~N.~Wong}
\affiliation{School of Natural Sciences, Institute for Advanced Study, 1 Einstein Drive, Princeton, NJ 08540, USA}
\affiliation{Princeton Gravity Initiative, Princeton University, Princeton, New Jersey 08544, USA}

\author[0000-0002- 7179-3816]{Daniel Palumbo}
\affiliation{Center for Astrophysics $\vert$ Harvard \& Smithsonian, 60 Garden Street, Cambridge, MA 02138, USA}
\affiliation{Black Hole Initiative at Harvard University, 20 Garden Street, Cambridge, MA 02138, USA}

\author[0000-0001-9939-5257]{Dominic Chang}
\affiliation{Black Hole Initiative at Harvard University, 20 Garden Street, Cambridge, MA 02138, USA}

\author[0000-0002-9031-0904]{Sheperd S. Doeleman}
\affiliation{Center for Astrophysics $\vert$ Harvard \& Smithsonian, 60 Garden Street, Cambridge, MA 02138, USA}
\affiliation{Black Hole Initiative at Harvard University, 20 Garden Street, Cambridge, MA 02138, USA}

\author[0000-0002-3351-760X]{Avery E. Broderick}
\affiliation{Perimeter Institute for Theoretical Physics, 31 Caroline Street North, Waterloo, ON, N2L 2Y5, Canada}
\affiliation{Department of Physics and Astronomy, University of Waterloo, 200 University Avenue West, Waterloo, ON, N2L 3G1, Canada}

\author[0000-0002-1919-2730]{Ramesh Narayan}
\affiliation{Center for Astrophysics $\vert$ Harvard \& Smithsonian, 60 Garden Street, Cambridge, MA 02138, USA}
\affiliation{Black Hole Initiative at Harvard University, 20 Garden Street, Cambridge, MA 02138, USA}

\author[0000-0002-8635-4242]{Maciek Wielgus}
\affiliation{Max-Planck-Institut f\"ur Radioastronomie, Auf dem H\"ugel 69, D-53121 Bonn, Germany}

\author[0000-0002-9030-642X]{Lindy Blackburn}
\affiliation{Center for Astrophysics $\vert$ Harvard \& Smithsonian, 60 Garden Street, Cambridge, MA 02138, USA}
\affiliation{Black Hole Initiative at Harvard University, 20 Garden Street, Cambridge, MA 02138, USA}

\author[0000-0002-0393-7734]{Ben S. Prather}
\affiliation{Department of Physics, University of Illinois, 1110 West Green Street, Urbana, IL 61801, USA}

\author[0000-0003-2966-6220]{Andrew A. Chael}
\affiliation{Princeton Gravity Initiative, Jadwin Hall, Princeton University, Princeton, NJ 08544, USA}
\affiliation{NASA Hubble Fellowship Program, Einstein Fellow}

\author[0000-0003-3457-7660]{Richard Anantua}
\affiliation{Department of Physics $\&$ Astronomy, The University of Texas at San Antonio, One UTSA Circle, San Antonio, TX 78249, USA}
\affiliation{Center for Astrophysics $\vert$ Harvard \& Smithsonian, 60 Garden Street, Cambridge, MA 02138, USA}
\affiliation{Black Hole Initiative at Harvard University, 20 Garden Street, Cambridge, MA 02138, USA}

\author[0000-0002-2825-3590]{Koushik Chatterjee}
\affiliation{Black Hole Initiative at Harvard University, 20 Garden Street, Cambridge, MA 02138, USA}

\author[0000-0003-3708-9611]{Ivan Marti-Vidal}
\affiliation{Departament d'Astronomia i Astrof\'isica, Universitat de Val\`encia, C/Dr. Moliner 50, 46100 Burjassot (Spain)}
\affiliation{Observatori Astron\`omic, Universitat de Val\`encia, C/Catedr\`atico Beltr\'an 2, 46980 Paterna (Spain)}

\author[0000-0003-4190-7613]{Jose L. Gómez}
\affiliation{Instituto de Astrofísica de Andalucía-CSIC, Glorieta de la Astronomía s/n}

\author[0000-0002-9475-4254]{Kazunori Akiyama}
\affiliation{Massachusetts Institute of Technology Haystack Observatory, 99 Millstone Road, Westford, MA 01886, USA}
\affiliation{Mizusawa VLBI Observatory, National Astronomical Observatory of Japan, 2-12 Hoshigaoka, Mizusawa, Oshu, Iwate 023-0861, Japan}
\affiliation{Black Hole Initiative at Harvard University, 20 Garden Street, Cambridge, MA 02138, USA}

\author[0000-0003-4475-9345]{Matthew Liska}
\affiliation{Center for Astrophysics $\vert$ Harvard \& Smithsonian, 60 Garden Street, Cambridge, MA 02138, USA}

\author[0000-0001-6950-1629]{Lars \ Hernquist}
\affiliation{Center for Astrophysics $\vert$ Harvard \& Smithsonian, 60 Garden Street,  Cambridge, MA 02138, USA}

\author[0000-0001-6950-1629]{Grant Tremblay}
\affiliation{Center for Astrophysics $\vert$ Harvard \& Smithsonian, 60 Garden Street,  Cambridge, MA 02138, USA}

\author[0000-0001-8593-7692]{Mark Vogelsberger}
\affiliation{Department of Physics, Kavli Institute for Astrophysics and Space Research, Massachusetts Institute of Technology, Cambridge, MA 02139, USA}

\author[0000-0002-7892-3636]{Charles Alcock}
\affiliation{Center for Astrophysics $\vert$ Harvard \& Smithsonian, 60 Garden Street, Cambridge, MA 02138, USA}

\author[0000-0003-4284-4167]{Randall  Smith}
\affiliation{Center for Astrophysics $\vert$ Harvard \& Smithsonian, 60 Garden Street, Cambridge, MA 02138, USA}

\author[0000-0002-5872-6061]{James Steiner}
\affiliation{Center for Astrophysics $\vert$ Harvard \& Smithsonian, 60 Garden Street, Cambridge, MA 02138, USA}

\author[0000-0003-3826-5648]{Paul Tiede}
\affiliation{Center for Astrophysics $\vert$ Harvard \& Smithsonian, 60 Garden Street, Cambridge, MA 02138, USA}
\affiliation{Black Hole Initiative at Harvard University, 20 Garden Street, Cambridge, MA 02138, USA}

\author[0000-0001-5461-3687]{Freek Roelofs}
\affiliation{Center for Astrophysics $\vert$ Harvard \& Smithsonian, 60 Garden Street, Cambridge, MA 02138, USA}
\affiliation{Black Hole Initiative at Harvard University, 20 Garden Street, Cambridge, MA 02138, USA}

\begin{abstract}
We investigate general relativistic magnetohydrodynamic simulations (GRMHD) to determine the physical origin of the twisty patterns of linear polarization seen in spatially resolved black hole images and explain their morphological dependence on black hole spin. By characterising the observed emission with a simple analytic ring model, we find that the twisty morphology is determined by the magnetic field structure in the emitting region. Moreover, the dependence of this twisty pattern on spin can be attributed to changes in the magnetic field geometry that occur due to the frame dragging. By studying an analytic ring model, we find that the roles of Doppler boosting and lensing are subdominant. Faraday rotation may cause a systematic shift in the linear polarization pattern, but we find that its impact is subdominant for models with strong magnetic fields and modest ion-to-electron temperature ratios. Models with weaker magnetic fields are much more strongly affected by Faraday rotation and have more complicated emission geometries than can be captured by a ring model. However, these models are currently disfavoured by the recent EHT observations of M87*. Our results suggest that linear polarization maps can provide a probe of the underlying magnetic field structure around a black hole, which may then be usable to indirectly infer black hole spins. The generality of these results should be tested with alternative codes, initial conditions, and plasma physics prescriptions. 

\end{abstract}


\section{Introduction}
The Event Horizon Telescope Collaboration (EHTC) has recently published the first polarized image of the supermassive black hole (SMBH) at the center of giant elliptical galaxy Messier 87 \citep[hereafter M87*;][]{PaperI,PaperII,PaperIII,PaperIV,PaperV,PaperVI,PaperVII,PaperVIII}. These results feature  
resolved linear polarization with a diffraction-limited resolution corresponding to approximately 5~$GM/c^2$, where $M$ is the mass of the SMBH, $G$ is the gravitational constant, and $c$ is the speed of light. The image reveals an asymmetric ring-like structure with a bright region at its Southern edge, attributed to the Doppler effect and the bending of light originating from the synchrotron emission of orbiting relativistic electrons in the vicinity of the event horizon. 

The polarimetric image of M87* BH 
presented in \citet{PaperVII, PaperVIII} provides information on 
both the degree and the directionality of linear polarization where the latter is determined by the electric vector position angle (hereafter EVPA). Quantifying the twistiness of the polarized images, defined as  the smooth azimuthal change of the EVPA within the ring pattern, offers us new insights into the structure of magnetic field, putting strong constraints on the nature of the ring and the emission region. 

The observed image of M87* from EHTC reveals an azimuthally spiraling pattern of the EVPA. Motivated by this, \cite{Palumbo_2020} used a particular decomposition of linear polarization to azimuthal modes identified with complex coefficients $\beta_m$. They found that the rotationally symmetric mode of a Fourier decomposition of a linear polarization pattern can distinguish between theoretical models. 
It was shown in \cite{PaperVIII} and \cite{Palumbo_2020} that $m=2$ (i.e. $\beta_2$) is the dominant contribution in the characterization of the magnetized accretion model. Furthermore, the $\beta_2$ phase matches very well with the theoretically expected behavior of the polarized map \citep[see][ for more details]{PaperVIII}. Finally, both the amplitude and the phase of $\beta_2$ are sensitive to the magnetic field geometry and the black hole spin. However, a detailed characterization of their correlation to the plasma astrophysics and the spacetime geometry remains elusive. 

\cite{Narayan_2021} modeled the polarimetric image of a BH using a simple toy model comprising of a magnetized ring of emission located near the Schwarzschild event horizon. The model comprises of an equatorial emission and fluid velocity with an arbitrary emission radius, magnetic field structure and observer's inclination. \citet{Gelles_2021} further extended this toy model to include the effect of BH spin by moving from the Schwarzschild spacetime to the Kerr geometry. The simplicity of the aforementioned 
toy models make them remarkably useful to do extensive
exploration of different emission models and magnetic field structure. Such an investigation is computationally very expensive using the general relativistic magnetohydrodynamical simulations (GRMHD). Yet, the validity of the assumptions made in such toy models remains elusive. Furthermore, it is not very clear what drives the $\beta_2$ in the simple ring model. 

Positron effects on polarized emission, including EVPA patterns, have been found to be dependent on plasma thermodynamics  in \cite{Anantua2020a,emami2021positron}. The linearly polarized portion of ray-traced images therein are supported on bilaterally asymmetric jet regions along with emitting rings around the Kerr magnetosphere. The global EVPA structure ranges from radial to spiral based on thermodynamic parameters and their interaction with the positron fraction of the plasma. 

In this work, we perform a comprehensive study of the driving sources of $\beta_2$ in simulated models of M87*. We utilize the same GRMHD simulations used in \citet{PaperV,PaperVIII} generated with the PATOKA pipeline \citep{2022ApJS..259...64W}, 
with a variety of different accretion states, BH spins and emission physics. To model the polarized images of M87*, we make use of the general relativistic radiative transfer (GRRT) framework implemented in code {\sc ipole} \citep{Moscibrodzka&Gammie2018}. We find the emission radius and link the magnetic field structure and velocity field at the emission location to the pattern of the EVPA. We investigate many effects which might contribute to the signal, including the spatial origin of the emission, Faraday rotation and relativistic effects. We find out that the signal strongly depends on the magnetic field geometry along with the BH spin. Our analysis show that while the ring model works reasonably good for the case of a magnetically arrested disk (MAD) \citep{1974Ap&SS..28...45B, 2003ApJ...592.1042I, 2003PASJ...55L..69N}, it has some limitations to fully capture the case of the standard and normal evolution (SANE) \citep{2003ApJ...599.1238D,GammieIHARM2003, 2012MNRAS.426.3241N}. Furthermore, the overall consistency reduces when we consider cooler electrons. We conclude that the phase of the $\beta_2$ indirectly probes the magnetic field geometry. Consequently, trends in $\beta_2$ are fundamentally linked to different magnetic field structures. 

The paper is organized as follows. Section~\ref{method} describes our methodology, including the GRMHD simulation~\ref{grmhd-sim}, the actual ray-tracing \ref{imaging-ipole}, magnetic field polarimetry \ref{B-field-polarimetry}, the geometrical ring model \ref{geometric-ring} and azimuthally expanded polarized mode ($\beta_2$) \ref{polarization}. Section~\ref{resul} focuses on the polarimetric analysis, including the time averaged polarized images \ref{time-averages}, calculation of the optical and the Faraday depths \ref{optical-depth},  the locus of emission \ref{emission-location}, the correlation between the phase of the $\beta_2$ and the BH spin \ref{phase-beta2-spin}, drivers of the $\beta_2$ \ref{driver-beta2}, the impact of the magnetic field on the phase of the $\beta_2$ \ref{B-V-spin-BH} and the influence of Faraday rotation \ref{FR-impact}. The conclusion is provided in Section~\ref{conclusion}. 

\section{Methodology} \label{method}
In this section, we describe how we produce the polarized images of M87*. More details about the simulation procedure and codes used can be found in \citet{2022ApJS..259...64W}.

\subsection{GRMHD Simulations} \label{grmhd-sim}
We use {\sc iharm} simulations by \cite{GammieIHARM2003,2021JOSS....6.3336P} from the standard library of 3D time-dependent GRMHD simulations performed in \citet{PaperV,PaperVIII}. These ideal GRMHD simulations are initialized with a weakly magnetized torus of plasma orbiting in the equatorial plane around a BH. Instabilities like the magnetorotational instability \citep{BalbusHowley1991} drive the torus into a turbulent state, which enables angular momentum transport and inward accretion of the matter onto the central black hole. The system tends toward a state with a mildly magnetized midplane, a coronal component where the gas to magnetic pressure $\beta \equiv P_g/P_B \simeq 1$ with a very strongly magnetized funnel region near the BH poles. The details of the outcome also depends on the strength and the geometry of the initial magnetic field. 

Due to the initial condition, the first part of each simulation is dominated by a transient state as a turbulent accretion develops . During this transient state, the accretion rate grows, and the infalling plasma heats up and begins to emit radiation. In order to ensure that this artificial initial transient state does not influence our results, we run each simulation until at least t =$ 10^4\,GM/c^3$, 
by which point the accretion flow close to the BH, which produces vast majority of the signal at $230\,$GHz, reaches a 
steady state. GRMHD fluid snapshots are saved every $5\,GM/c^2$
over the duration of the simulation; the time range of the initial transient state is found by analyzing the fluid snapshot data so that the ``steady-state'' epoch of the simulation can be identified. 

For GRMHD simulations with non-zero BH angular momentum, we study the cases where the BH angular momentum $J$ is aligned (parallel or anti-parallel) with the angular momentum of the accretion flow. 
Ideal GRMHD simulations are invariant under mass rescaling and thus our (anti-)aligned systems are effectively described by just two parameters: the BH angular momentum (spin) and the near horizon magnetic flux $\Phi_B$. The dimensionless BH spin, hereafter $a \equiv Jc/GM^2$, is limited to $ -1 \leqslant a \leqslant 1$. The dimensionless magnetic flux at the horizon is $\phi \equiv \Phi \left( \dot{M} r^2_g c \right)^{-1/2}$ (where $r_g \equiv GM/c^2$ refers to the gravitational radius) determines if the accretion is in a SANE ($\phi \simeq 5 $) or MAD ($\phi \gtrsim $50) state.  

The numerical methods used to solve the MHD equations often fail in the regions with low density $\rho$ and strong magnetic fields $B$. These failures are dealt with through the application of floors, which inject artificial density or energy, producing unreliable plasma temperatures. To ensure that the artificial floors do not influence simulated observations, we employ a $\sigma$ cutoff during the radiative transfer calculation and mask regions where $\sigma = B^2/\rho > 1$, following \citet{PaperV}. Typically the impact on SANE simulations is negligible, as the flagged region is concentrated around the axis, where very little emission is predicted. In MAD models the masked region may occur in different locations of the simulation, and it is expected to result in slightly dimmer mm wavelength images. While this problem is fundamental and important to study further, we expect that it should not appreciably impact the image morphology, and proceed under this assumption. In this work, we adopt $\sigma = 1$ as the boundary between the regions with reliable plasma parameters and the aforementioned flagged region.
To illustrate the possible impact of the $\sigma$ in our analysis, in Figures \ref{MAD-Emission} and \ref{SANE-Emission}
we present the contour of $\sigma =1$ with dash-dot white line. Furthermore, in Figure \ref{Sigma-R-Z-plot} we present the color-plot of the $\log_{10}{\sigma}$ in the source $R$ vs $Z$ plane for the time and azimuthally averaged GRMHD simulations of MAD and SANE simulations. The 
emission location is marked as the black star. It is generally seen than the the emission location is fairly consistent with the white region, associated with $\sigma =1$ as adopted in this work.

\subsection{Imaging} \label{imaging-ipole}
In order to generate polarized images of M87*, we use the GRRT code {\sc ipole} \citep{Moscibrodzka&Gammie2018}. Each image is produced with a 160 $\mu$as field of view (FOV) and a resolution of 320 $\times$ 320 pixels, with each image containing the 4 Stokes parameters, I, Q, U and V. {\sc ipole} first solves for the null geodesic equations from the camera through the source, then the polarized radiative transfer equations forward along with the geodesic. Polarized synchrotron emission, self-absorption, Faraday rotation, and Faraday conversion are also taken into account in making the polarized images. For each set of model parameters (see below), images are produced over the entire steady-state time range identified in the GRMHD simulations.

Unlike in the GRMHD simulations, the GRRT calculation, which relies on the emission, absorption, and rotation transfer coefficients, is not scale-invariant. Thus, while performing the ray-tracing, we must set two physical scales for the system. The first scale is the characteristic system length, which is computed as $L = GM/c^2$. 
The second scale is determined by the simulation mass-density parameter, which is set by the observed flux density at 230 GHz which is chosen to be $F_{\nu} \simeq 0.5$ Jy \citep{PaperIV}. Following \citet{PaperV}, we have also fixed the inclination of the source to be 17 deg for retrograde, and 163 deg for the prograde spins. 

Next, to set the plasma temperature, we choose a different strategy than the one usually adapted in GRMHD simulations. In particular, we replace the thermal equilibrium approximation with a collisionless plasma in which electrons and ions most likely reach two different temperatures \citep{Shapiro+1976,Narayan+1995}.  Hence, as in previous work \citep[e.g.,][]{PaperV,PaperVIII}, we modulate the ion-to-electron temperature ratio via the prescription of \citet{Monika+2016}:
\begin{equation}
\frac{T_i}{T_e} = R_\mathrm{high} \frac{\beta^2}{1+\beta^2} + R_\mathrm{low} \frac{1}{1+\beta^2}.
\end{equation}
Here, $\beta$ is the ratio of gas to magnetic pressure, and $R_\mathrm{high}$ and $R_\mathrm{low}$ are free parameters, allowed to vary with guidance from simulations that include electron heating.  In our ``Fiducial'' set of models, we set $R_\mathrm{low}=1$ and $R_\mathrm{high}=20$, values favored by recent simulations \citep{Chael+2018,Mizuno+2021}. In our ``Faraday Thick'' set of models, on the other hand, we set $R_\mathrm{low}=10$ and $R_\mathrm{high}=160$, considered large values for each parameter. The electrons are colder in the Faraday Thick models, requiring increased mass density to match the observed flux of M87*.  Both the lower temperature and the increased density makes this model much thicker to Faraday rotation than the Fiducial case, which plays an important role in decreasing the linear polarization fraction and modifying the EVPAs of a given region. A significant variation of the Faraday depth toward M87* has been reported by \citet{Goddi2021}, motivating us to address both regimes.

The version of {\sc ipole} that we use has been modified to record the spatial distribution of the emission that contributes to an image. Notice that this value depends on the location of the observer, due to both absorption (which occurs as light travels along the geodesic) and the anistropy of synchrotron emission (which makes the relevant emission a function of the line of sight through any particular points in space). In practice, we compute the observed emission for each pixel in an image and then sum over the pixels. For the geodesic corresponding to any image pixel, it is trivial to compute the effects of absorption along the geodesic by evaluating the optical depth between each point along that pixel's geodesic and the camera at infinity. The local (angle-dependent) emissivity can be directly computed at each event along the geodesic and multiplied into the optical depth extinction to calculate the local contribution to the final observed flux density. After repeating this procedure for each image's pixel, a histogram of all local contributions to the final image can be computed. This histogram represents the origin of all observed emission. More detail can be found in \S~3.2.1 of \cite{2022ApJS..259...64W}.

\subsection{Magnetic Field Polarity} \label{B-field-polarimetry}
The equations of ideal GRMHD are invariant under a sign flip of the magnetic field direction. Previous EHTC studies have considered only the case where the overall magnetic field polarity is parallel to that of the disk angular momentum. As we shall show however, this choice impacts the linear polarization structure in the images because of the Faraday rotation which is in place when the linear polarization passes through a magnetized plasma. Consequently, in generating images, we consider two distinct cases: one set with the magnetic field vector aligned with the disk angular momentum and the other is anti-aligned. Below, we examine both of these cases, calling them FR$_1$ and FR$_2$, respectively. 

\subsection{Geometric Ring Model} \label{geometric-ring}
To build intuition for the polarized images of the M87* accretion flow produced by the EHT \citep{PaperVIII}, \citet{Narayan_2021} constructed a toy model for the synchroton emission from a thin ring of magnetized fluid orbiting a Schwarzschild black hole. This model assumes an optically thin ring of fluid at a single Boyer-Lindquist radius with an axisymmetric magnetic field and fluid velocity. Predictions for the observed polarization pattern on the viewing screen take the shape of a lensed ring of polarization vectors. 

\citet{Gelles_2021} extended the ring model to address equatorial emission in a Kerr spacetime and also specified the emission from highly lensed sub-images for which geodesics complete one or more half-orbits around the black hole. The Gelles model found that, assuming all other fluid parameters are equivalent, the effect of spin on the direct image polarization is very weak; therefore, the difference in polarization patterns that trend with spin are being driven indirectly, likely through modification of the underlying velocity and magnetic field directions, as we will show.

As shown in the GRMHD comparison paper by \citet{Narayan_2021}, these toy models can be used to produce non-infinitesimal rings by evaluating them over a range of radii and applying an envelope function to the underlying emissivity. \citet{Palumbo_KerrBAM} applied this procedure to the \citet{Gelles_2021} model in Kerr metric to produce the image generation and model-fitting code ``Kerr Bayesian Accretion Modeling'' (\texttt{KerrBAM}), which we use to generate toy model images throughout the paper. \texttt{KerrBAM} generates model image by semi-analytically ray-tracing backwards from the observer screen to an arbitrarily inclined equatorial plane in Boyer-Lindquist coordinate, producing grids of radii and azimuthal angles. Given an axisymmetric prescription for the fluid velocity and magnetic field penetrating the plane, in addition to an envelope function $\mathcal{J}$, the synchrotron emissivity can be predicted for a given spectral index, which we take it to be 1 at 230 GHz (see the discussion in subsection 2.2 of \citet{Narayan_2021}). Throughout this paper, we use an axisymmetric ring profile given by:
\begin{align}
    \label{eqn:profile}
    \mathcal{J}_{\rm ring}(r) &= \exp{\left(-4 \ln2 \frac{(r-R)^2}{w^2}\right)},
\end{align}
where $R$ is the peak radius and $w$ is the full width of half maximum of the intensity profile. Note that the true emissivity also depends on the details of lensing, Doppler boosting, and the angle between geodesics and the local magnetic field, all of which are accounted for in \texttt{KerrBAM} for arbitrary photon winding numbers. However, the model does not contain any Faraday effects and thus does not meaningfully predict fractional polarization. Consequently, the ring model is most useful in the context of our paper as a predictor of the resolved structure of the EVPA. 

\begin{figure*}[th!]
\center
\includegraphics[width=0.99\textwidth]{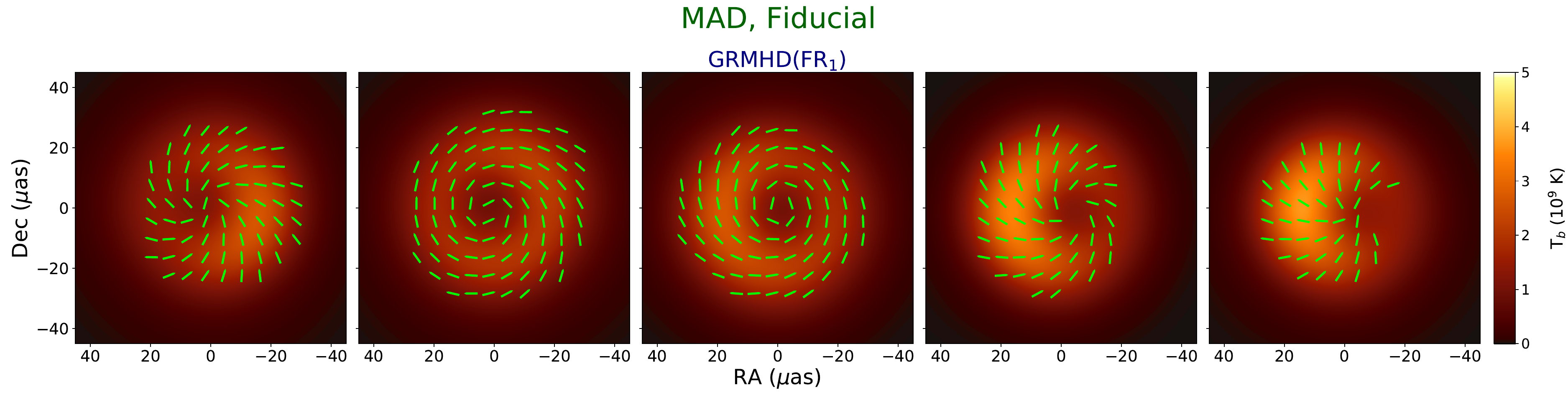}
\includegraphics[width=0.99\textwidth]{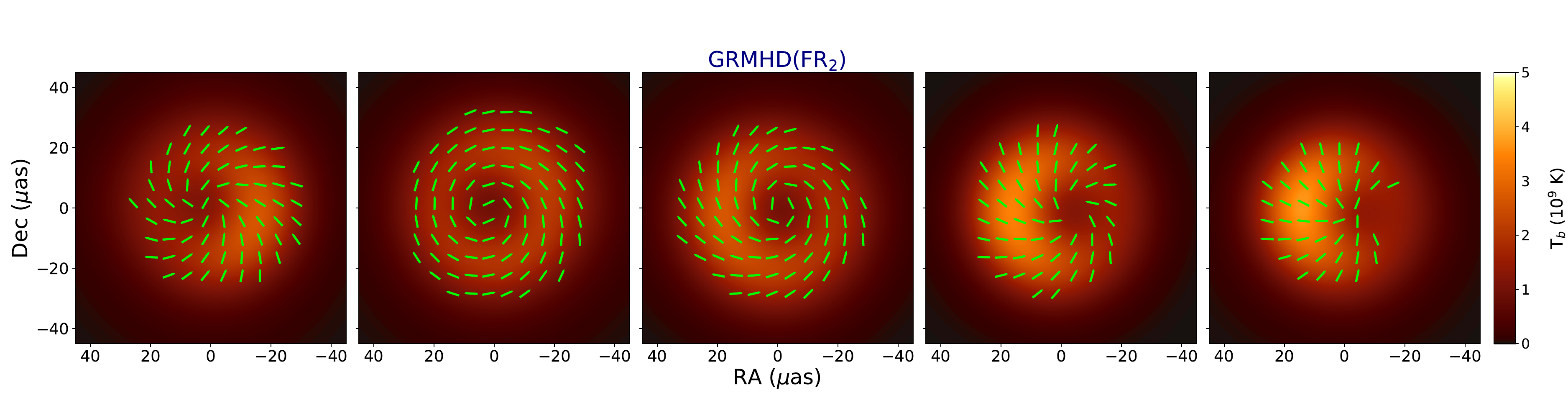}
\includegraphics[width=0.99\textwidth]{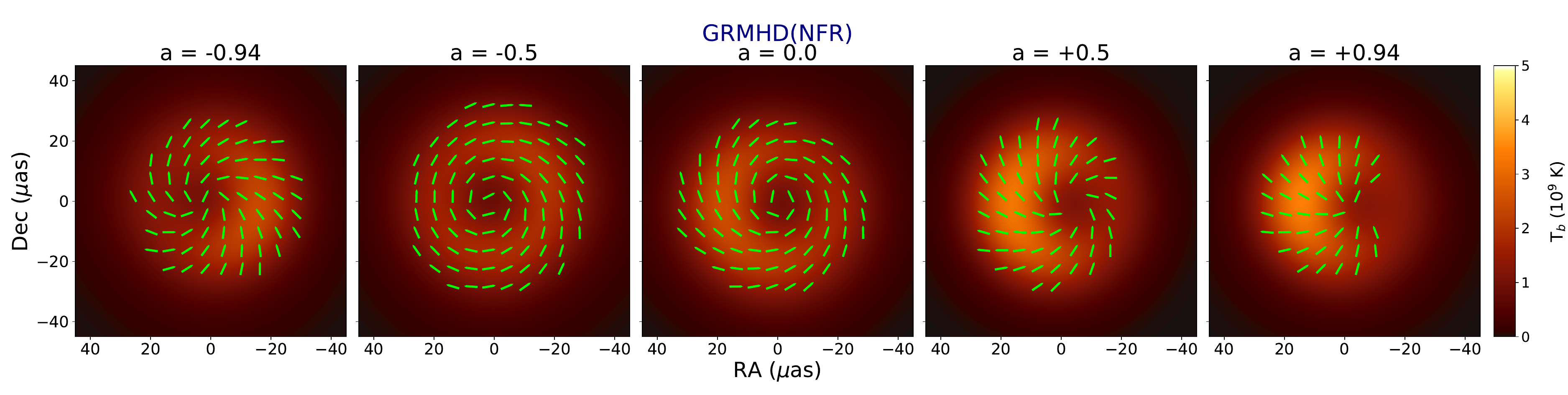}
\includegraphics[width=0.99\textwidth]{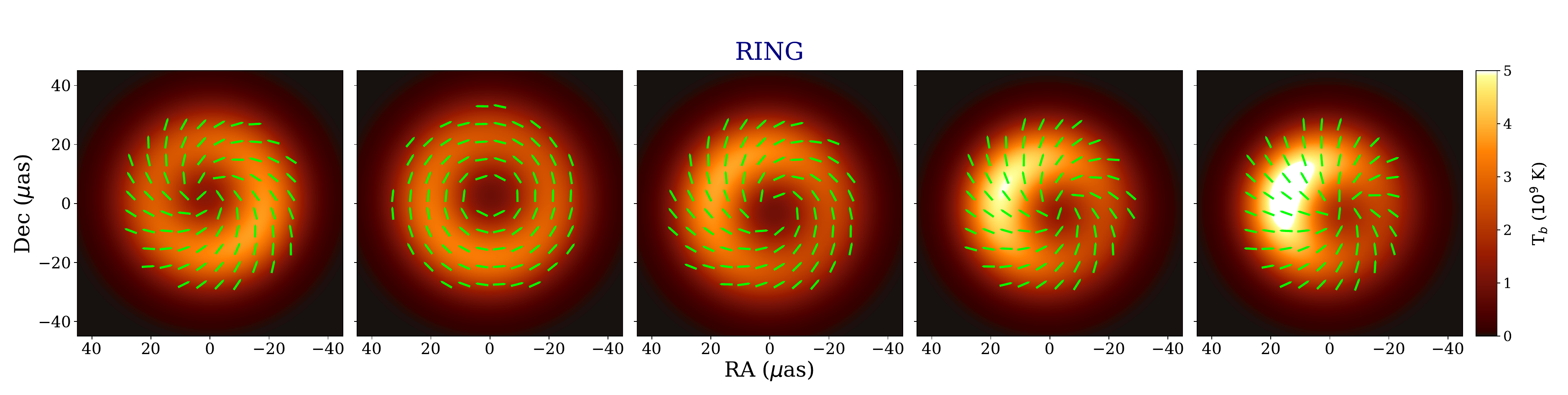}
\caption{Time averaged images of MAD models for the Fiducial case with $R_{\mathrm{high}} = 20$ and $R_\mathrm{low}=1$ with different BH spins, $a=(-0.94, -0.5, 0.0, +0.5, +0.94)$. The first two rows plot polarimetric images computed using the complete radiative transfer equation, with two orientations of the magnetic field polarity, either aligned with the disk angular momentum on large scales (FR$_1$) or anti-aligned (FR$_2$).  In the third row, we switch off Faraday rotation in this calculation to study its impact on the linear polarization pattern, which is not very significant for these models.  Finally, in the fourth row, we plot images derived from an analytic ring model with parameters chosen to match the GRMHD. This ring model reproduces the evolution of the twisty morphology as a function of spin by including evolution in the magnetic field structure, velocity field, and emission location (but not Faraday rotation) as input from GRMHD.}
\label{Mad-averaged-fiducial}
\end{figure*}

\begin{figure*}[th!]
\center
\includegraphics[width=0.99\textwidth]{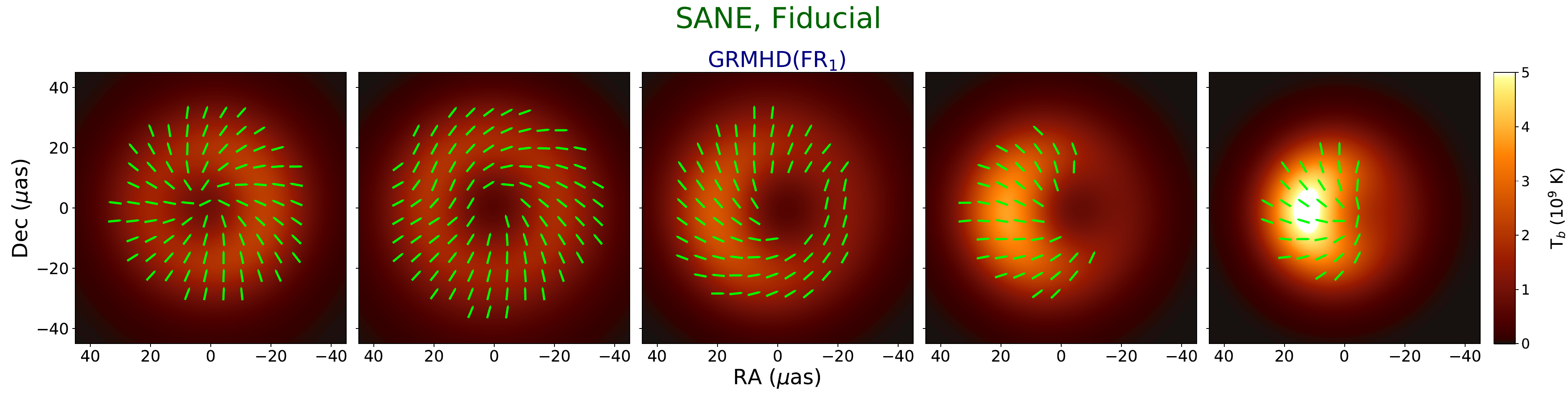}
\includegraphics[width=0.99\textwidth]{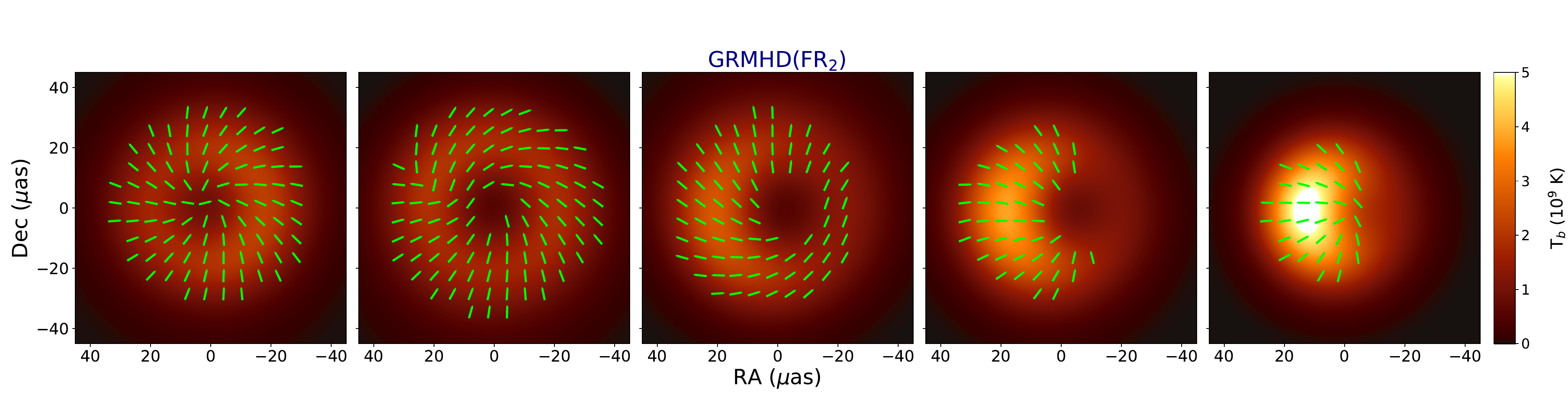}
\includegraphics[width=0.99\textwidth]{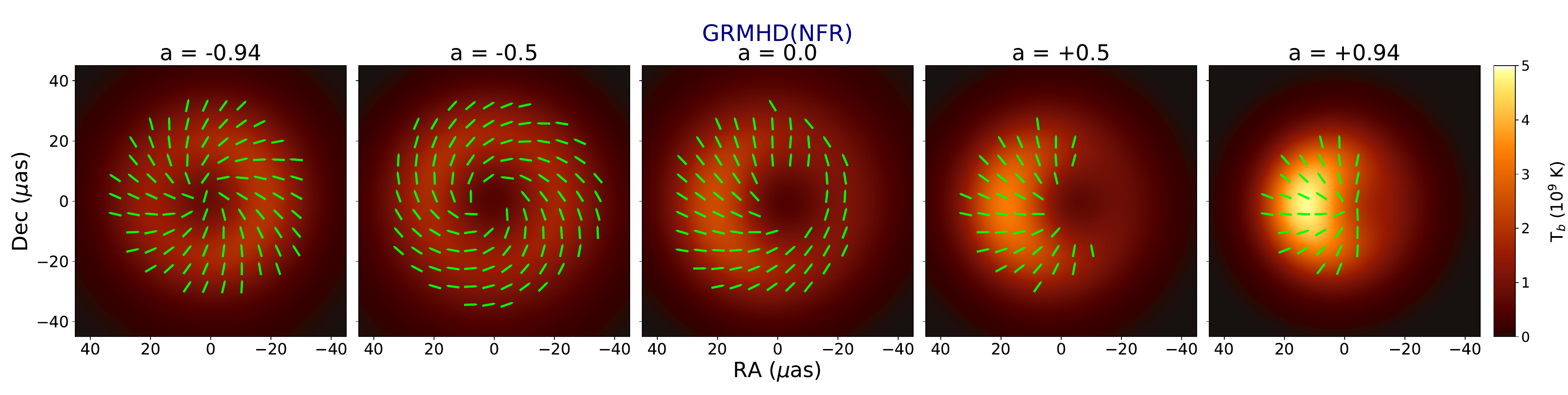}
\includegraphics[width=0.99\textwidth]{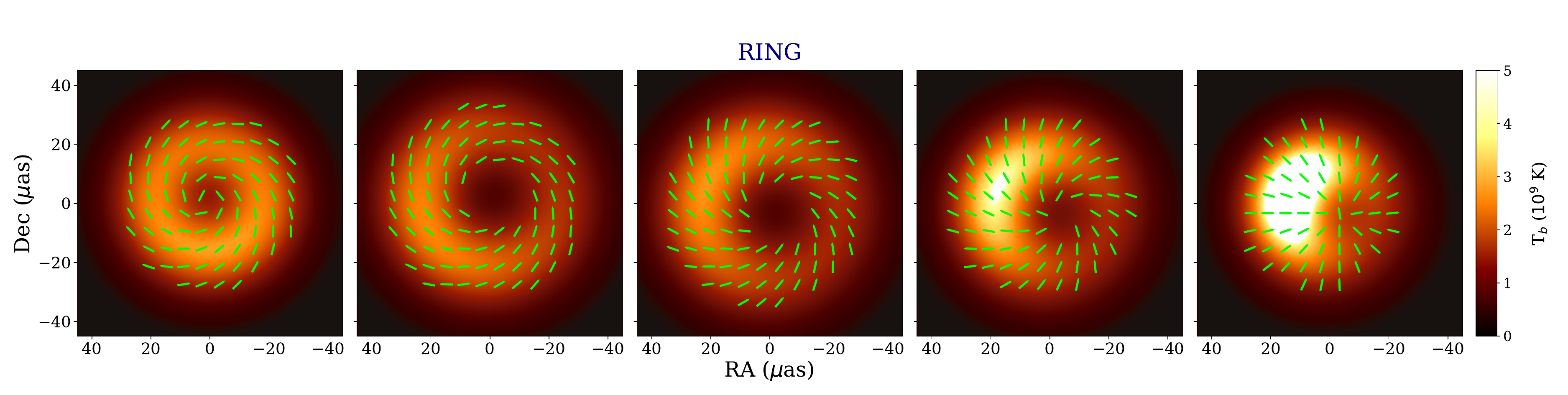}
\caption{As \autoref{Mad-averaged-fiducial}, but for our Fiducial SANE models.  The impact of Faraday rotation is more significant due to their larger Faraday depths.  While a ring model can broadly reproduce the nature of the morphology evolution as a function of spin, more deviations occur for our SANE models due to their more complicated emission geometries. \label{Sane-averaged-fiducial}}
\end{figure*}
\subsection{Polarization Spirals and \texorpdfstring{$\beta_2$}{beta-2}} \label{polarization}
\citet{Palumbo_2020} found that the magnetization state in simulations of the M87* accretion flow was strongly encoded in the spiraling structure of the EVPA, with more radially directed EVPA corresponding the SANE models and more spiraling or circular EVPA patterns corresponding to MADs. To differentiate between the MAD and SANE models more quantitatively, they proposed an azimuthal decomposition of the polarized image to isolate symmetric patterns in the EVPA with respect to the image azimuthal angle $\varphi$, when averaged over image radius $\rho$:
\begin{align}
    \beta_m &=\dfrac{1}{I_{\rm tot}} \int\limits_{0}^{\infty} \int\limits_0^{2 \pi} P(\rho, \varphi) \, e^{- i m \varphi} \; \rho \mathop{\ud\varphi}  \mathop{\ud\rho},\\
    I_{\rm tot} &= \int\limits_{0}^{\infty} \int\limits_0^{2 \pi} I(\rho, \varphi) \; \rho \mathop{\ud\varphi} \mathop{\ud\rho}.
\end{align}
where $I$ refers to the the intensity while $P = Q + i U$ describes the linear polarization with $Q$ and $U$ referring to the Stokes parameters.
Here, each coefficient $\beta_m$, which is normalized by the total image flux, is a complex coefficient with a phase that encodes the orientation of the spiral and an amplitude that is proportional to the average linear fractional polarization. Crucially, the $m=2$ mode is rotationally symmetric; intuitively, a nearly axisymmetric flow viewed at nearly face-on inclination (as is the case for M87*) would produce a nearly rotationally symmetric image on the observer screen. \citet{Palumbo_2020} found that, as expected, the $\beta_2$ mode is indeed the most informative coefficient for discriminating the accretion states, as the phase directly relates to the underlying magnetic field orientation in the limit of weak Faraday effects. Furthermore, \citet{PaperVIII} found that the phase of $\beta_2$ was more constraining than any other polarized image metric when comparing images from EHT data to GRMHD simulations. Of interest in this work is how strongly the $\beta_2$ phase can be related to the magnetic field in GRMHD simulations in two regimes of Faraday effect strength, as well as investigating the cause of the strong dependence on spin in the $\beta_2$ phase identified in \citet{Palumbo_2020}.

As in \cite{PaperVIII}, we use the default image centering of the GRMHD library (which corresponds to a zero angular momentum photon at the screen center) when evaluating $\beta_2$, rather than recentering based on an empirical ring search. As evaluated in the discussion of \citet{Palumbo_2020}, the amplitude of $\beta_2$ decays quadratically in the centering error (as a fraction of the ring diameter), while phase effects enter only at higher order. In highly asymmetric images, the selection of an appropriate image center in observations may be more complicated, involving image-domain feature extraction; one might also choose the image center that maximizes the $\beta_2$ amplitude.

\subsection{From GRMHD simulation to the geometrical Ring model} \label{connection}
Having introduced the GRMHD models and the geometrical ring model, here we briefly discuss on how to make the connection between the two. This is essential as the ring model chooses an arbitrary value for the magnetic and the velocity fields. Consequently, to make a direct link between the $\beta_2$ phase from the GRMHD and the ring model, we shall make use of the B/V fields from the GRMHD simulations which requires some coordinate transformations. More appropriately, while the GRMHD snapshots are given in the Kerr-Schild coordinate, the Ring model is based on the Boyer-Lindquist coordinate.

\begin{figure*}[th!]
\center
\includegraphics[width=0.99\textwidth]{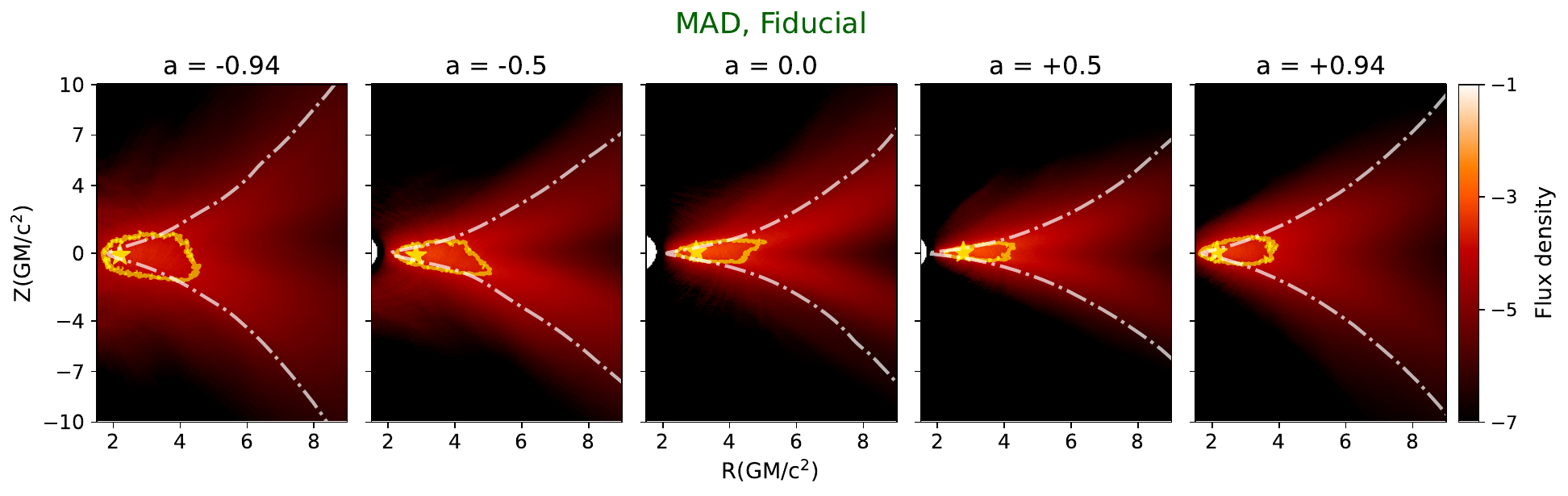}
\includegraphics[width=0.99\textwidth]{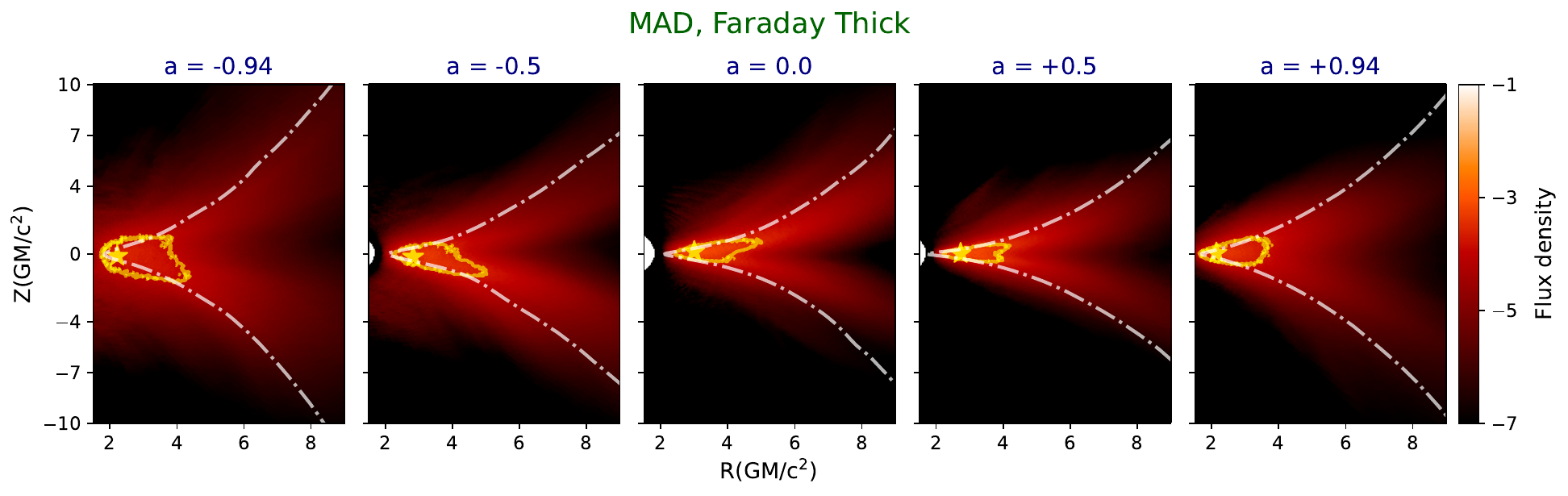}
\caption{Azimuthal and time-averaged emissivity densities for our Fiducial (top) and Faraday Thick (bottom) MAD simulations.  The yellow contour surrounds the region within which the top 30\% of the emission is localized, while the gold star marks the single chosen emission location used for extracting the magnetic and velocity fields for the ring model.  Dash-dot white lines indicate the $\sigma=1$ contour. In each row, from the left to right we increase the BH spin in the interval $a=(-0.94, -0.5, 0.0, +0.5, +0.94)$. Our MAD models, even those with large $R_\mathrm{high}$ and $R_\mathrm{low}$ are characterized by a ring of emission in the mid-plane.}
 \label{MAD-Emission}
\end{figure*}

Comparison between GRMHD and the ring model requires establishing a suitable mapping that relates a given simulation to a ring model product. We created this mapping for each simulation by first selecting the radius of emission in the ring model to be the same as the peak emission radius of the chosen simulation, (see \autoref{MAD-Emission}). We then azimuthally extracted the equatorial magnetic fields and fluid velocities from the simulation. 
\newline A series of coordinate transformations were performed to convert the GRMHD output into a suitable ring model input. The fluid velocity is reported in a local inertial frame oriented along the FMKS coordinate axis, (see Appendix F of \cite{2022ApJS..259...64W} for a definition), while the magnetic field is extracted from GRMHD as a 3-vector in the fluid frame. These quantities were used to construct a four-velocity and magnetic field four-vector as defined in \cite{GammieIHARM2003}, which were transformed to Boyer-Lindquist coordinates. 
Finally, we constructed the necessary four vector quantities for the ring model by transforming from Boyer-Lindquist to the equatorial ZAMO frame with the appropriate tetrads defined in \cite{Gelles_2021}, and performing the required Lorentz boost to retrieve the magnetic field vector in the fluid frame.

\section{Polarimetric Analysis and results} \label{resul}

In order to determine the origin of the linear polarization structure in our images, we perform an in-depth analysis of the GRMHD simulations at the location of peak emission.  First, we locate the emissivity peak and sample the magnetic fields and velocity fields there.  We input this information in an analytic ring model that treats Doppler boosting and lensing self-consistently (but does not include Faraday effects).  We use this methodology to identify the mechanisms that determine $\angle \beta_2$ which describes the argument of $\beta_2$.

\subsection{Time averaged polarized images} \label{time-averages}
Throughout our analysis, we mainly focus on the time averaged images from different snapshots of a GRMHD simulations. That is, we are assuming that the turbulent character of the flow is stationary, and the source morphology may be characterized with a well-defined mean image. This notion appears to be consistent with the results of the multi-year M87* total intensity monitoring \citep{Wielgus2020}. Figures \ref{Mad-averaged-fiducial} and \ref{Sane-averaged-fiducial} present the time averaged images of the GRMHD simulations and the geometrical ring model for the Fiducial case of MAD and SANE simulations, respectively. In each figure, the first two rows present images with aligned and anti-aligned magnetic fields (referred to as FR$_1$ and FR$_2$ respectively), followed by the case with no Faraday rotation (referred to as NFR) and the ring model (named as RING).  Since the ring model does not include Faraday rotation, we are mostly concerned with comparing the NFR row with the RING row.  In each row, from the left to right, we increase the BH spin in the interval $a=(-0.94, -0.5, 0.0, +0.5, +0.94)$. 

From these plots, it is evident that the linear polarization pattern is not strongly affected by Faraday rotation. This itself is an important finding, since Faraday rotation and depolarization could have potentially randomized $\angle \beta_2$ in our models.  Moreover, despite the abundance of simplifying assumptions made in constructing the ring models, they are remarkably successful at reproducing the overall handedness of the twisty polarization pattern especially in MAD simulations. SANE models, on the contrary, exhibit less agreement with the simple ring model. Consequently, the linear polarization pattern in MAD simulations can be explained by the magnetic field and velocity field at the emission peak, as input parameters in ring model, while its spin evolution can be attributed to frame dragging. This is explored in much greater detail in the following sections.

The azimuthally oriented EVPAs in low BH spins in MAD simulations are getting converted to radial ticks for high BH spins (both retrograde and the prograde spins) with more radial EVPAs for the prograde spins. Consequently, it is inferred that the ticks of the EVPAs might be directly linked to the BH spin and can be used to infer the spin. The azimuthal/radial pattern in MAD is getting slightly distorted in SANEs owing to the extra scrambling induced from the Faraday rotation. The radially oriented EVPAs in SANE prograde spins is also reflected in the ring model, though it is absent in the retrograde spins. 

Since the electron-to-ion temperature ratio remains an important uncertainty in our models, this need not generically be the case.  As constructed, Faraday rotation has a more significant effect in our Faraday Thick models.  Their images are discussed in more detail in Appendix \ref{Image-Faraday-Thick}.

\begin{figure*}[th!]
\center
\includegraphics[width=0.99\textwidth]{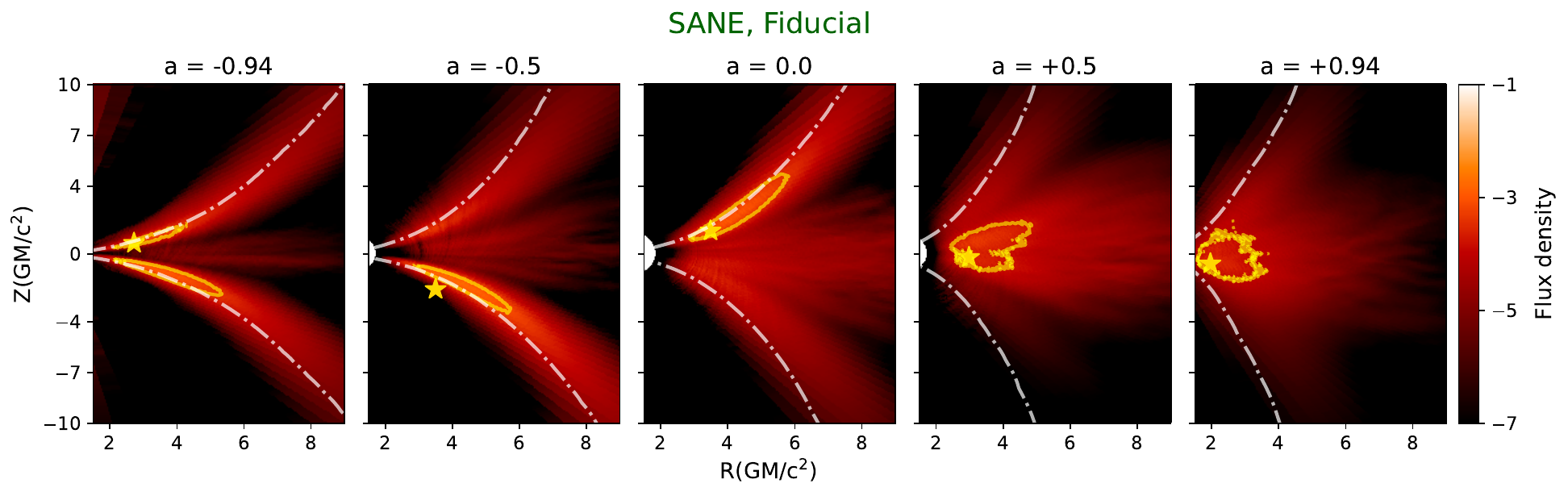}
\includegraphics[width=0.99\textwidth]{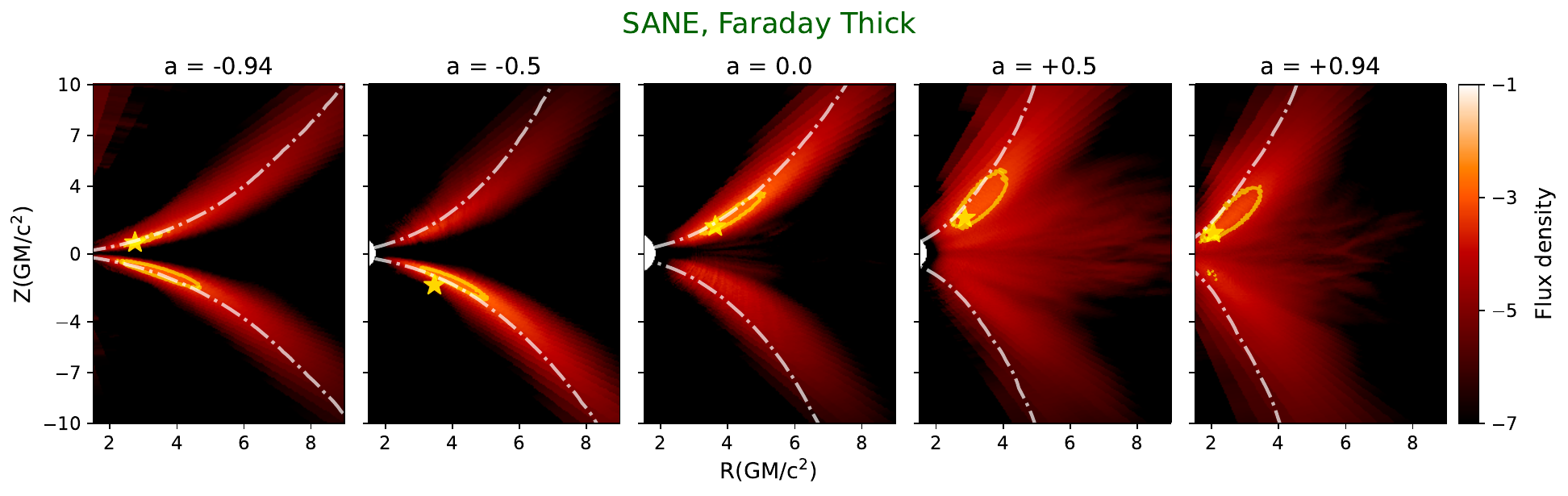}
\caption{As \autoref{MAD-Emission}, but for our SANE simulations.  Our SANE models have more complicated emission geometries and in some cases clearly exhibit multiple peaks of emission.  Since we pick a single emission location to input into our ring model, we expect greater disagreement between GRMHD and analytic rings for our SANE models than their MAD counterparts.  In each row, from the left to right we increase the BH spin in the interval $a=(-0.94, -0.5, 0.0, +0.5, +0.94)$.}
 \label{SANE-Emission}
\end{figure*}


\begin{figure*}[th!]
\center
\includegraphics[width=0.99\textwidth]{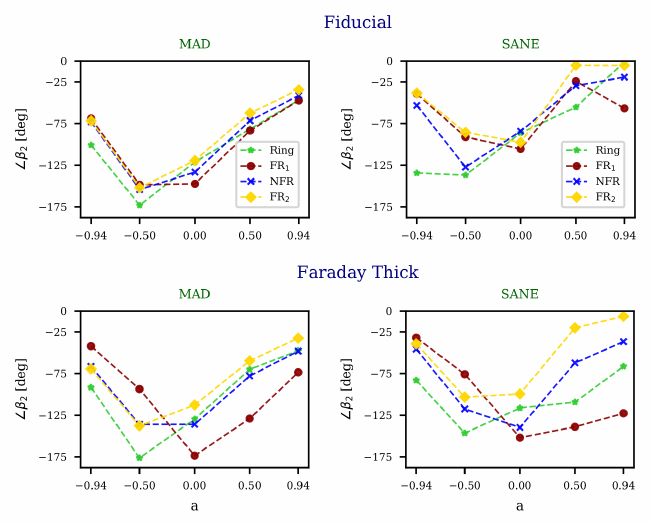}
\caption{$\angle \beta_2$ for different GRMHD simulations and the ring model. FR$_1$, FR$_2$, NFR and the ring model are marked with red-circle, yellow-diamond, blue-cross and green-stars, respectively.  The top row plots our Fiducial models, while the bottom row plots our Faraday Thick models.  The left column plots our MAD models, while the right column plots our SANE models.  A simple analytic ring given the appropriate velocity and magnetic fields does a remarkable job at reproducing $\angle \beta_2$ of our Fiducial MAD models.  In other models, the differences between the NFR and FR$_{1,2}$ points reveal a stronger impact of Faraday rotation.
 }
\label{Phase_Comparison}
\end{figure*}

\subsection{Inferring the optical depth and Faraday depth for polarized images} 
\label{optical-depth}
In the following, we infer the optical depth and the Faraday depth for images with different electron temperature profiles.  Tables \ref{Table-OD-FD-MAD} and \ref{Table-OD-FD-SANE} present the mean and the standard deviation of the optical depth (OD) and the Faraday depth (FD) for the time-averaged GRMHD simulations for the Fiducial (Fid) and the Faraday thick (Thick) cases with $FR_1$ and $FR_2$ models for MAD and SANE simulations with $a = (-0.94, -0.5, 0.0, +0.5, +0.94)$, respectively. It is inferred from the tables that both of the OP and the FD are higher for the Faraday Thick case than the Fiducial case. Furthermore, while the FD is prominently higher in SANE simulations compared with the MADs, the OD is almost the same between MAD and SANEs for the retrogrades while it is higher in SANEs for the prograde cases.

\begin{table*}[th!]
\caption{The time averaged of optical depth (OD) and the Faraday depth (FD) for the Fiducial (Fid)  and the Faraday thick (Thick) cases for $FR_1$ and $FR_2$ cases for MAD simulations with $a = (-0.94, -0.5, 0.0, +0.5, +0.94)$.}
\begin{flushleft}
\begin{tabular}{l|cc|cc|cc|cc|ccr}
\toprule
\hline
{\color{blue} MAD} & \multicolumn{2}{c}{a = -0.94} &  \multicolumn{2}{c}{a = -0.5} & \multicolumn{2}{c}{a = 0.0} & \multicolumn{2}{c}{a = +0.5} & \multicolumn{2}{c}{a = +0.94}  \\ \hline
models & {\color{blue} FD}  & {\color{red} OD}  & {\color{blue} FD} & {\color{red} OD} & {\color{blue} FD} & {\color{red} OD} & {\color{blue} FD} & {\color{red} OD} & {\color{blue} FD} & {\color{red} OD}  \\
\hline 
Fid ($FR_1$) &  2.1 $\pm$ 0.4 & 0.07 $\pm$ 0.03 & 2.6 $\pm$ 0.9 & 0.07 $\pm$ 0.03  &  2.1 $\pm$ 0.6 & 0.09  $\pm$ 0.03 &  1.7 $\pm$ 0.5 & 0.13 $\pm$ 0.04 &  0.9 $\pm$ 0.2 & 0.09 $\pm$ 0.04  \\ \hline 
Fid ($FR_2$) &  2.1 $\pm$ 0.4 & 0.07 $\pm$ 0.03 & 2.6 $\pm$ 0.9 & 0.07 $\pm$ 0.03 & 2.1 $\pm$ 0.6 & 0.09  $\pm$ 0.03 & 1.7 $\pm$ 0.5 & 0.13 $\pm$ 0.04 &
0.9 $\pm$ 0.2 & 0.09 $\pm$ 0.04  \\ \hline 
Thick ($FR_1$) &  360 $\pm$ 71 & 0.83 $\pm$ 0.33 &  433 $\pm$ 102 & 0.76 $\pm$ 0.28 &  386 $\pm$ 81 & 0.82 $\pm$ 0.30 & 405 $\pm$ 125  & 1.34 $\pm$ 0.45  & 176 $\pm$ 58 & 1.06 $\pm$ 0.40  \\ \hline 
Thick ($FR_2$) &  360 $\pm$ 71 & 0.83 $\pm$ 0.33 &  433 $\pm$ 102 & 0.77 $\pm$ 0.28 &  387 $\pm$ 81 & 0.83 $\pm$ 0.34 & 406 $\pm$ 126  & 1.35 $\pm$ 0.45  & 176 $\pm$ 58 & 1.07 $\pm$ 0.40        \\ \hline 
\bottomrule 
\end{tabular}
\end{flushleft}
\label{Table-OD-FD-MAD}
\end{table*}

\begin{table*}[th!]
\caption{The time averaged of optical depth (OD) and the Faraday depth (FD) for the Fiducial (Fid) and the Faraday thick (Thick) cases for $FR_1$ and $FR_2$ cases for SANE simulations with $a = (-0.94, -0.5, 0.0, +0.5, +0.94)$. }
\begin{flushleft}
\begin{tabular}{l|cc|cc|cc|cc|ccr}
\toprule
\hline
{\color{blue} SANE} & \multicolumn{2}{c}{a = -0.94} &  \multicolumn{2}{c}{a = -0.5} & \multicolumn{2}{c}{a = 0.0} & \multicolumn{2}{c}{a = +0.5} & \multicolumn{2}{c}{a = +0.94}  \\ \hline
models & {\color{blue} FD($\times 10^3 $)}  & {\color{red} OD}  & {\color{blue} FD($\times 10^3 $)} & {\color{red} OD} & {\color{blue} FD($\times 10^3 $)} & {\color{red} OD} & {\color{blue} FD($\times 10^3 $)} & {\color{red} OD} & {\color{blue} FD($\times 10^3 $)} & {\color{red} OD}  \\ \hline 
Fid ($FR_1$) &  8.2 $\pm$ 2.5 & 0.05 $\pm$ 0.03 & 17.8 $\pm$ 3.0 & 0.06 $\pm$ 0.03 & 69 $\pm$ 47 & 1.0 $\pm$ 1.6 & 2.1 $\pm$ 0.5 & 1.4 $\pm$ 0.9 & 0.2 $\pm$ 0.05 & 1.1 $\pm$ 0.9   \\ \hline 
Fid ($FR_2$) &  8.2 $\pm$ 2.5 & 0.05 $\pm$ 0.03 & 17.8 $\pm$ 3.0 & 0.07 $\pm$ 0.03 & 69 $\pm$ 47 & 1.0 $\pm$ 1.7 & 2.2 $\pm$ 0.5 & 1.5 $\pm$ 0.9 & 0.2 $\pm$ 0.05 & 1.1 $\pm$ 0.9  \\ \hline 
Thick ($FR_1$) &  267 $\pm$ 88 &  0.38 $\pm$ 0.15 &  657 $\pm$ 101 & 0.28 $\pm$ 0.08 &  4114 $\pm$ 2789 & 0.45 $\pm$ 0.11  &  1790 $\pm$ 425 & 6.0 $\pm$ 4.9 &  140 $\pm$ 46.8 & 9.1 $\pm$ 12.0  \\ \hline 
Thick ($FR_2$) & 268 $\pm$ 88 &  0.38 $\pm$ 0.15 &  656 $\pm$ 101 & 0.28 $\pm$ 0.08 &  4128 $\pm$ 2798 & 0.45 $\pm$ 0.12  &  1878 $\pm$ 428 & 6.7 $\pm$ 5.6 &  140 $\pm$ 47 & 9.3 $\pm$ 12.3  \\ \hline 
\bottomrule 
\end{tabular}
\end{flushleft}
\label{Table-OD-FD-SANE}
\end{table*}

\subsection{Quantifying the emission location in GRMHD simulations} 
\label{emission-location}
As described in Section \ref{imaging-ipole}, we construct for each snapshot a three-dimensional map of the emission location, which is then averaged over the time and azimuth to identify the $r$ and $z$ coordinates of the emission peak.  In Figures \ref{MAD-Emission} and \ref{SANE-Emission}, we plot the azimuthal and the time averages (see Section \ref{time-averages} for more details). Specifically, we identify the region containing the top 30\% of the emission, outlined in white. Marked with gold star, we also present the chosen emission location as appropriate in the ring model.
\cite{2022ApJ...931...25T, 2022Galax..10..103T} showed that the dominant emission location differs between the total intensity, the linear and the circular polarization. More specifically that the total intensity is mostly originated from the jet base while the linear polarization comes from the downstream of the jet and the circular polarization is maximum at the counter-jet. In our analysis, we use a mixed method, where we identify the top 30\% of emission based on the intensity and mark them in Figs.~\ref{MAD-Emission}-\ref{SANE-Emission}, while we fix the emission location, indicated with a star in Figs.~\ref{MAD-Emission}-\ref{SANE-Emission}, such that the inferred $\angle \beta_2$ from the ring model gets as close as possible to the results of the GRMHD models. 

We find that the emission of our Fiducial MAD models is concentrated in the mid-plane.  Meanwhile, SANE simulations reveal a more complicated emission geometry \citep[see also][]{PaperV}. In the Fiducial case, the emission of prograde models is mostly located in the mid plane, but emission becomes more jet dominated in the other cases.  Several of the SANE models are clearly not well described by a single ring.  In the Faraday Thick SANE case, disk emission is further suppressed, moving more emission into the jet sheath.  In conclusion, while the emission location is robust in MADs against changing the electron temperature, it depend on the details of the electron distribution in SANE simulations.

\subsection{\texorpdfstring{$\angle \beta_2$}{The Angle of beta-2} as a Function of Spin} \label{phase-beta2-spin}

\begin{figure*}[th!]
\center
\includegraphics[width=0.99\textwidth]{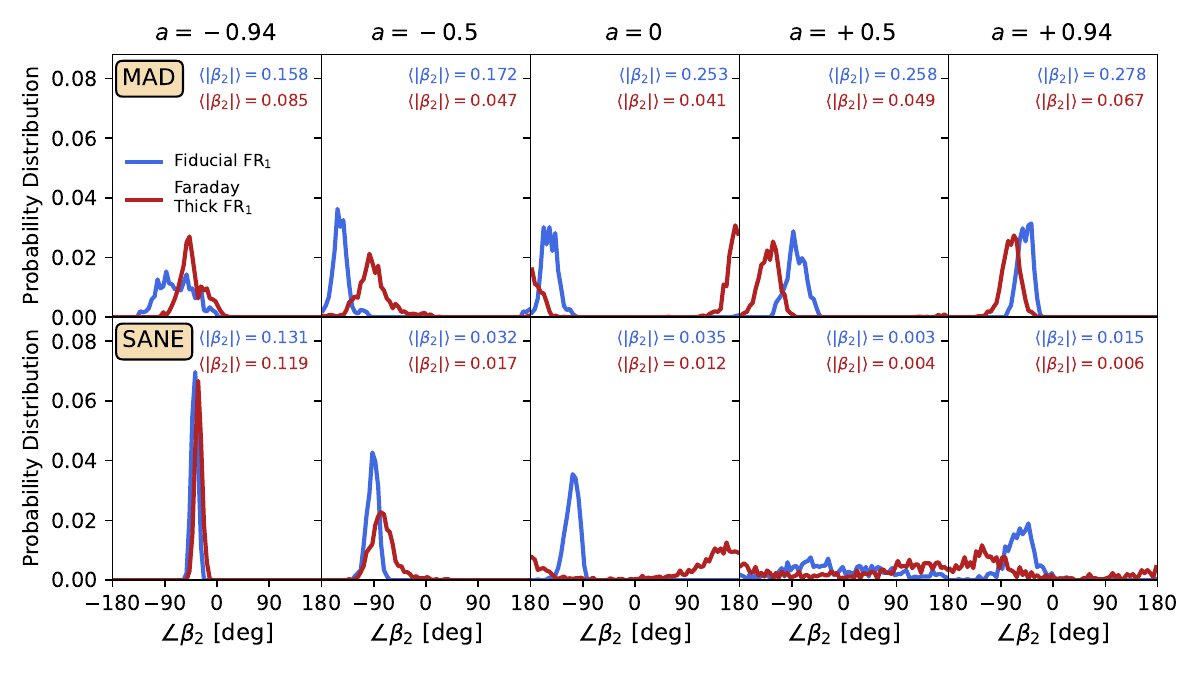}
\caption{Distributions of $\angle \beta_2$ in our models.  Fiducial models are shown in blue while Faraday Thick models are shown in red.  At the top of each panel, we write the average amplitude of the $\beta_2$ mode as well.  For models with $\langle |\beta_2| \rangle \gtrsim 0.1$, we find that these peaks are generally well-localized.  For some SANE models with scrambled $|\beta_2|$, $\angle \beta_2$ is randomized and carries little information.}
 \label{fig:beta_distributions}
\end{figure*}

In analyzing distributions of the complex $\beta$ coefficients of GRMHD simulations, \citet{Palumbo_2020} point out an interesting dependence of $\angle \beta_2$ on the SMBH spin.  We reproduce this result and plot both the Fiducial and Faraday Thick models (with magnetic fields aligned with the disk angular momentum; FR$_1$) in \autoref{fig:beta_distributions}.  The differences in these distributions between aligned and anti-aligned magnetic fields are discussed in \autoref{sec:flip_b}.  We find that most models exhibit relatively localized peaks of $\angle\beta_2$, even among our Faraday Thick models. Fortunately, Faraday rotation does {\it not} randomize $\angle \beta_2$ for a given snapshot.  We find that our prograde SANE models are poorly localized however.  This is because they exhibit very low $|\beta_2|$ as written at the top right of each panel, and thus their $\angle \beta_2$ is not very meaningful.  As discussed in \citet{PaperVIII}, even though the retrograde SANEs exhibit a high Faraday depth, this Faraday depth does not affect forward-jet emission that is in front of the Faraday screen.  Finally, while outside the scope of this paper, it is clear that the width of these distributions varies in interesting ways among these different models, which can be constrained by continued monitoring of M87*.

Here we compute the $\angle \beta_2$ in various GRMHD simulations, comparing them against the geometrical ring model. From the GRMHD part, we infer the phase of $\beta_2$ for FR$_1$, FR$_2$ and NFR, while from the ring side, we compute the $\angle \beta_2$ by choosing few different locations with high percentage of emission, fixing the magnetic and velocity fields in one final location where the inferred $\angle \beta_2$ from the ring model is the closest to the NFR case. The chosen place is then marked with gold stars in Figures \ref{MAD-Emission} and \ref{SANE-Emission}. 

\begin{figure*}[th!]
\center
\includegraphics[width=0.99\textwidth]{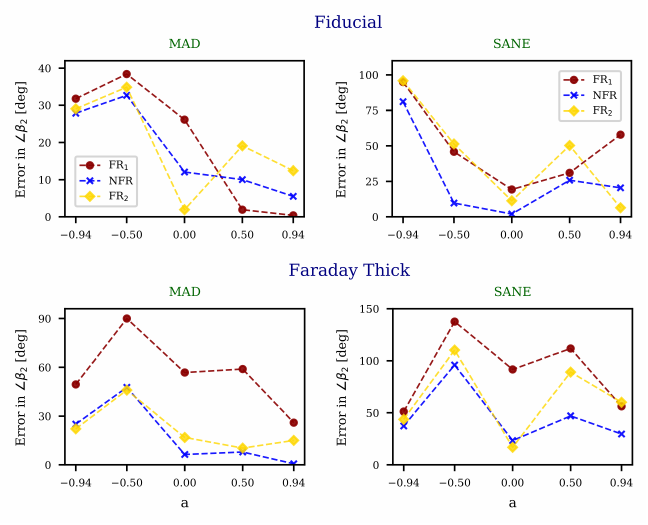}
\caption{The error in the $\angle \beta_2$ for different GRMHD simulations compared with the ring model. FR$_1$, FR$_2$ and NFR are marked with red-circle, yellow-diamond and blue-cross , respectively. The top/bottom panels show the Fiducial/Faraday Thick cases, respectively. The left(right) panel show the MAD(SANE) simulations.  Since Faraday rotation is not included in our simple ring model, it naturally produces the smallest errors when compared to the NFR cases. 
}
\label{Ring-Comparison}
\end{figure*}

Figure \ref{Phase_Comparison} and \ref{Ring-Comparison} present the $\angle \beta_2$ and the Error in $\angle \beta_2$ in different GRMHD simulations and the ring model, respectively. In the top panel, we present the Fiducial case while in the bottom, we show the Faraday Thick case. In each row, the left/right panel, presents the MAD/SANE case. From the plots, it is inferred that:

$\bullet$ Overall, MAD simulations establish better agreement with the ring model than the SANE simulations.

$\bullet$ The level of the model agreements is higher in the Fiducial case compared with the Faraday Thick scenario. Furthermore, FR$_1$ and FR$_2$ are more similar in the former case than the latter one. This is expected as the Faraday rotation is more prominent in the Faraday Thick case than the Fiducial one.

$\bullet$ In the Faraday Thick case, the anti-aligned magnetic field, FR$_2$, gets closer to the NFR case than the aligned magnetic field. 

$\bullet$ Finally, it is seen that the $\angle \beta_2$ is directly linked to the BH spin. Consequently, we argue that the $\angle \beta_2$ in MAD and SANE simulations can be used to infer the BH spin. 

\subsection{Main drivers of \texorpdfstring{$\beta_2$}{beta-2} } \label{driver-beta2}
Here we explore different drivers of the ticks in EVPAs. The most key drivers of the EVPAs include plasma boost, gravitational lensing and the magnetic field geometry. While the first two players are easier to probe individually, we may not turn off the magnetic field as it washes out the $\beta_2$ entirely. Owing to this, in what follows, we directly explore the impact of the plasma boost as well as the gravitational lensing, while indirectly estimating the importance of the magnetic field based on the deviations of EVPA ticks from their original values when we turn off the boost or the gravitational lensing. 

\subsubsection{Influence of plasma boost on \texorpdfstring{$\beta_2$}{beta-2}}
\begin{figure*}[th!]
\center
\includegraphics[width=0.99\textwidth]{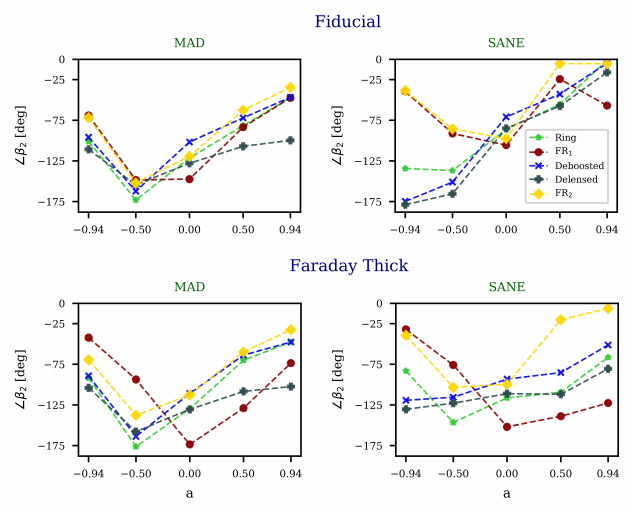}
\caption{The impact of the plasma boost and the lensing on the phase of \texorpdfstring{$\beta_2$}{beta-2}. Doppler boosting and lensing both have mild effects on $\angle \beta_2$, but do not drive its evolution with spin.  }
\label{Ring-Deboosting-Delensing}
\end{figure*}

The main role of the plasma boost factor can be traced by turning off the $\beta$ in the ring model. The blue-crossed-dashed lines in figure \ref{Ring-Deboosting-Delensing} presents the $\angle \beta_2$ in the absence of the boost factor. From the plot, it is inferred that in MAD simulations the deboosted lines generally follow the main trends in the $\angle \beta_2$, as is shown with green-star-dashed lines.  SANEs on the other hand deviate slightly from the main trend at low-retrograde spins, while they show a more reasonable behavior at higher spin cases. 
In conclusion, while the plasma boost is one of the players in driving the $\angle \beta_2$, it is not the key driver! 

\subsubsection{Impact of gravitational lensing on \texorpdfstring{$\beta_2$}{beta-2}}
Next, we analyze the influence of the gravitational lensing on the EVPA ticks by scaling the BH mass down by a large factor, in our case by a factor of 1000, and scaling up all of coordinates by the same factor in order to preserve the angular size of the image. By placing emission at large radii in gravitational units, the impact of lensing is removed completely, altering the observed intensity and polarization by changing the angle of emission of geodesics that ultimately land at the observer screen. The impact of lensing is necessarily greatest in high-inclination models, while here we consider only nearly face-on models. Among these low inclination models, these changes will have greater impact in models with large fluid velocities (where Doppler effects increase angular dependence of emitted intensity) and in models with larger variation in $\vec{k} \times \vec{B}$ across the image. 

The plus-grey-dashed line in Figure \ref{Ring-Deboosting-Delensing} illustrates the impact of delensing on  $\angle \beta_2$. From the plot one can infer that, in MAD retrograde and low prograde spins, the delensed lines lie very close to the full ring models while the high spin prograde cases show a substantial deviation from the original ring model. Furthermore, the delensed cases in SANE simulations follow the same trends as in deboosted ones meaning that both of these effects are not the main drivers in $\angle \beta_2$. 

\subsubsection{Influence of magnetic field geometry on the \texorpdfstring{$\beta_2$}{beta-2}}
\label{B-V-spin-BH}

\begin{figure*}[th!]
\center
\includegraphics[width=0.99\textwidth]{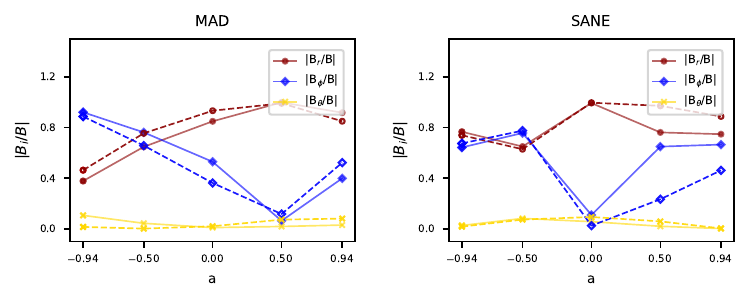}
\includegraphics[width=0.99\textwidth]{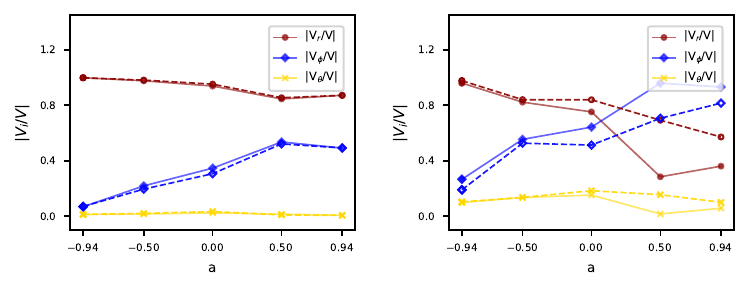}
\caption{(Top) the $|B_{i}/B|$ vs the BH spin, (bottom) $|V_{i}/V|$ as a function of the BH spin for MAD vs the SANE simulation at the location of the emission for $R_{\mathrm{high}}=20$ (filled-solid lines) as well as $R_{\mathrm{high}}=160$ (empty-dashed lines). Evidently, the $r$ and $\phi$ field's components are dominant over the $\theta$ components for MAD simulations and are nearly the case for SANE simulations as well. }
\label{B-V-field-amp}
\end{figure*}

\begin{figure*}[th!]
\center
\includegraphics[width=0.99\textwidth]{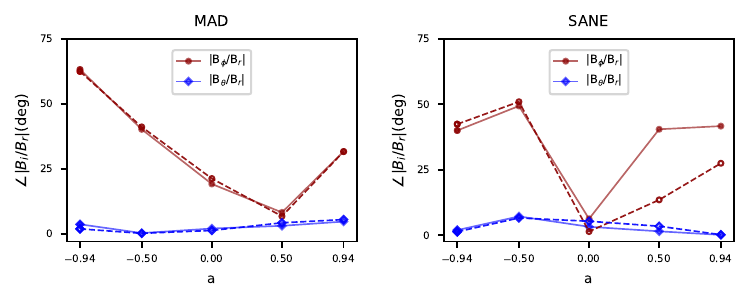}
\includegraphics[width=0.99\textwidth]{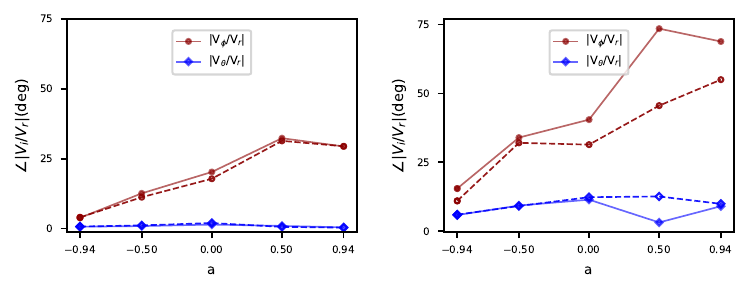}
\caption{(Top) the $\angle{(|B_{i}/B_{r}|)} \equiv \arctan{(|B_{i}/B_{r}|)}$
vs the BH spin, (bottom) $\angle{(|V_{i}/V_{r}|)} \equiv \arctan{(|V_{i}/V_{r}|)}$  with $i = (\phi, \theta)$ as a function of the BH spin for MAD vs the SANE simulation at the location of the emission. }
\label{B-V-field-angle}
\end{figure*}

Having found that boosting and lensing play relatively minor roles in driving $\angle \beta_2$, we now investigate the magnetic and the velocity fields at the emission location to elucidate trends with the BH spin.  Figure \ref{B-V-field-amp} presents the absolute value of the normalized components of the magnetic field (top row) as well as the velocity field (bottom row) for our Fiducial model (solid-lines) as well as the Faraday Thick case (dashed-lines). From the plot, it is evident that the radial and azimuthal components of the B and V fields show a clear trend with the BH spin.  In comparison, the polar components (vertical, when sampling the midplane) remain insignificant in comparison at the peak of the emitting region.  Furthermore, in MAD simulations, this trend is not sensitive at all to the electron temperature, while in SANE simulations there is a very little dependence on the electron temperature in prograde spins.  Note that since the magnetic field and velocity field are GRMHD primitives, any change as a function of electron temperature corresponds to a change in the peak of the emitting region.  Consequently, we conclude that the combination of the B and V fields at the emission location might be very informative in constraining the BH spin.

Figure \ref{B-V-field-angle} presents the $\angle{(|B_{i}/B_{r}|)}$ (top row) and $\angle{(|V_{i}/V_{r}|)}$ (bottom row) with $i = (\theta, \phi)$ as a function of the BH spin. In each row, the left(right) panel shows the MAD(SANE) models. The solid (dashed) lines show the Fiducial (Faraday thick) cases. There is a clear trend of the aforementioned quantities with the BH spin. The trend is clearer for the azimuthal component than the polar one. Both MAD and SANE simulations show similar behavior for each of the above quantities. For instance, in both cases $\angle{(|B_{\phi}/B_{r}|)}$ has a turn over behavior in which the angle is first decreasing and then is enhancing in the prograde regime, though the exact turn over point is not the same in MAD and SANE simulations. $\angle{(|V_{\phi}/V_{r}|)}$, on the other hand, establishes an enhancing pattern in both cases, though the slope of the enhancement is not the same in MAD and SANE. 

These trends are naturally explained in the context of frame dragging of magnetic field lines.  A generic helical pattern develops in disk-fed accretion flows from flux-freezing and frame dragging \citep[e.g.,][]{Semenov+2004,Ricarte+2021}.  First, we find that higher prograde spin values result in higher azimuthal velocities, as expected.  This becomes reflected in the image as more Doppler boosting and thus greater asymmetry.  More importantly, the magnetic field grows more tangential as spin increases, in either the prograde or retrograde cases. The exact ratio should depend on exactly the radius at which the emission peaks. \citet{Ricarte+2022} explicitly showed the evolution of these components as a function of radius in a similar set of simulations, and in retrograde systems, a sign flip in this ratio can result in a similar sign flip in $\angle \beta_2$ with radius.

\subsection{Impact of Faraday Rotation} \label{FR-impact}

As linear polarization travels through a magnetized plasma, Faraday rotation shifts the EVPA by an amount depending on the intervening density, temperature, and magnetic field.  This effect could significantly impact $\angle \beta_2$ for two reasons.  First, Faraday rotation imprints the line-of-sight direction of the magnetic field, rotating ticks counterclockwise if the field is pointed towards the observer and clockwise if the field is pointed away, which directly impacts $\angle \beta_2$.  Second, large Faraday depths can lead to depolarization/scrambling, both along the line-of-sight and between neighboring regions.  Our analytic ring model does not incorporate Faraday rotation, and thus it is important to determine to what extent it can impact our results.

\begin{figure*}[th!]
\center
\includegraphics[width=0.99\textwidth]{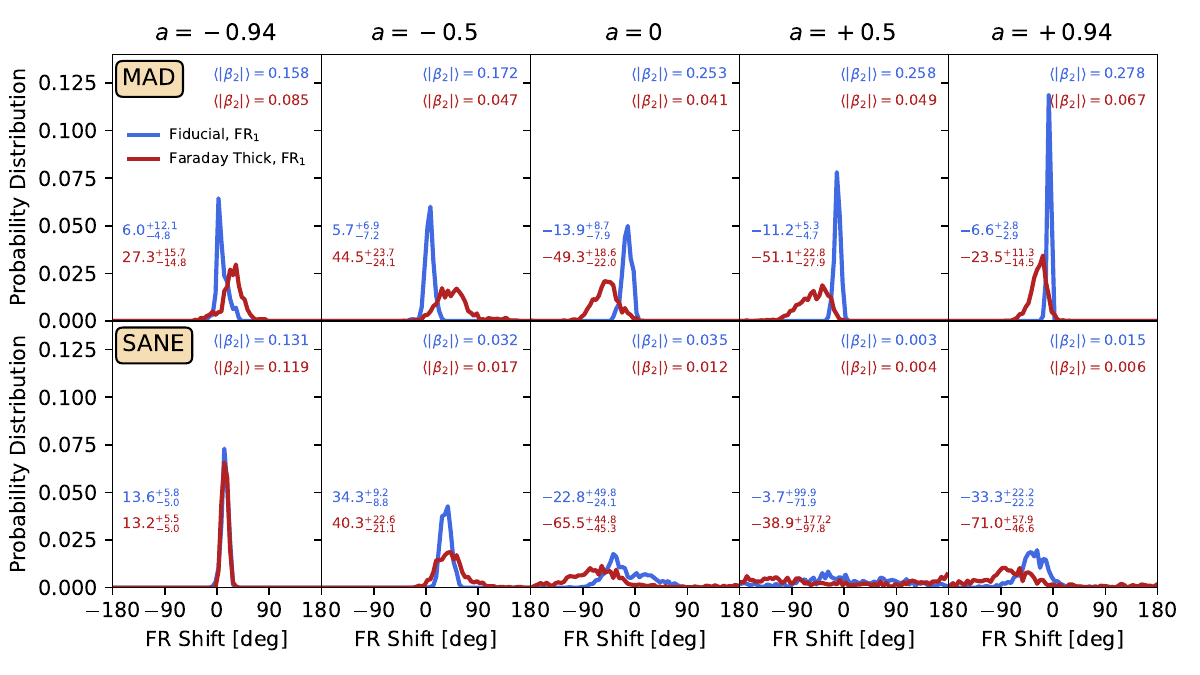}
\caption{For each of the models, we compute the shift in $\angle \beta_2$ due to Faraday rotation and here plot the probability distribution of these shifts.  At the top of each panel, we also write the average amplitude of $\beta_2$ to show how important the mode is for each model.  The text in each panel quotes 16th, 50th, and 84th percentiles for these shifts.  For models where $\langle | \beta_2 | \rangle$ is significant ($\gtrsim 0.1$), we find that these shifts are modest, demonstrating that Faraday rotation typically only has a small impact on $\beta_2$ for these models.  However, Faraday rotation can cause a much greater shift if SANEs and our Faraday Thick models.  We find an interesting bias induced by the assumed direction of the magnetic field.  Retrograde models are preferentially shifted towards more positive values, while prograde (and spin 0) models are shifted towards more negative values due to the assumption that the vertical magnetic field is aligned with the outer disk angular momentum, which is reversed by construction in our retrograde models.}
 \label{fig:beta_offset_20_1}
\end{figure*}

To directly assess the affects of Faraday rotation, we compute images with Faraday rotation switched off (setting the coefficient $\rho_V=0$) for each model.  After computing $\angle \beta_2$ for images with Faraday rotation switched off and comparing with the images with Faraday rotation switched on as normal, we plot the distribution of shifts in $\angle \beta_2$ induced by Faraday rotation in \autoref{fig:beta_offset_20_1}.  
The blue lines show the Fiducial models, while the red lines present the Faraday Thick cases. For most models in the Fiducial set (and all of the MADs), we find that the impact of Faraday rotation is small, typically shifting $\angle \beta_2$ by roughly 10 degrees.  For most of the SANEs in the Fiducial set, this effect can be more significant.  However, much of this can be explained by the fact that $|\beta_2|$ is low (written on the top right of each panel), and thus is not carrying much information to begin with.

We find an interesting pattern in the shifts due to Faraday rotation that differs between the prograde and retrograde models: retrograde models are preferentially shifted toward more positive values of $\angle \beta_2$ while prograde models are preferentially shifted toward more negative values.  This pattern is consistent with a bias induced by the assumed direction of the vertical magnetic field, aligned here with the disk angular momentum on large scales. This is discussed in more detail in \autoref{sec:flip_b}. 

In summary, we find that Faraday rotation fortunately usually does not randomize $\angle \beta_2$, even in some severely Faraday Thick models.  However, it can impart a systematic shift in the distributions. This shift is small, approximately $10^\circ$ for our Fiducial MAD models, but can potentially be much more significant for Faraday Thick and SANE models.  We find an interesting signature of the polarity of the magnetic field that we discuss further in \autoref{sec:flip_b}.

In the set of Fiducial simulations considered in this paper, the poloidal field direction is arbitrarily assumed to be parallel with the BH spin.  The field direction could equivalently have been oriented in the opposite direction without affecting the evolution of the GRMHD.  Consequently, any image library including only poloidal fields aligned with the BH spin vector is incomplete. 

\section{Conclusions} \label{conclusion}
We have performed an in-depth exploration of the origin of the 
``twistiness'' of the EVPAs, quantified by the $\angle \beta_2$, in simulated, time-averaged polarized images of M87*. We used a simple ring model to take into account different contributions including the plasma Doppler boosting, gravitational lensing and magnetic field geometry, in driving the ticks of EVPAs. We inferred the main location of the emission in various GRMHD simulations and read the magnetic and velocity fields at the emission location as key ingredients in our geometrical ring model. Our results can be summarized as follows:  

$\bullet$ Comparing the geometric ring model with the GRMHD simulations, it is inferred that MAD models in general provide a better agreement with the ring model than the SANE cases.

$\bullet$ The Fiducial case shows a higher level of agreement than the Faraday Thick models, as expected, owing to the fact that the Faraday rotation is more important in the latter case than the former one. 

$\bullet$ $\angle \beta_2$ seems to be directly linked to the BH spin, therefore it may be possible to use the twistiness of the ticks of the linear polarization to the infer the BH spin.

$\bullet$ Our analysis showed that among the drivers of the ticks of the EVPAs, the plasma boost and gravitational lensing provide a sub-dominant contribution in determining $\angle \beta_2$. Consequently, the magnetic field geometry predominantly drives the ticks of the linear polarization. 

$\bullet$ We have shown that there is a trend in the normalized amplitude of the radial and azimuthal components of the magnetic and velocity fields at the emission location with the BH spin. This is encouraging, as any future measurement of the magnetic field and the plasma velocity might be very informative about the BH spin. Consequently, we propose to use the next generation of the EHT (ngEHT) to look for the BH spin through a measurement of the B and V fields at the emission location. 

$\bullet$ We found that the impact of Faraday rotation on $\angle \beta_2$ is small for our Fiducial models. However, we noticed an interesting bias induced by the alignment vs anti-aligned of the vertical magnetic field direction with the BH spin.  This bias becomes much more significant in the Faraday Thick models, which have relatively cooler electrons and increased Faraday depths compared to the Fiducial case. Future studies using libraries of model images from GRMHD should also include models where the magnetic field direction is flipped, which fortunately does not require additional GRMHD simulations.

$\bullet$ Although we considered a wide range of models in this paper, we did not explore the impact of different initial conditions on trends in $\angle \beta_2$. Higher resolutions could also qualitatively alter the flow when the plasmoids form. Finally, different emission models including non-thermal models, positrons and the tilted disk models may also contribute in changing the results. We leave further explorations of the above cases to a future study. 

\section*{Data Availability}
Data directly corresponded to this manuscript and the figures is available to be shared on reasonable request from the corresponding author. The ray tracing of the simulation done in this work was performed using the  {\sc ipole} \citep{Moscibrodzka&Gammie2018}. We have used the library of {\sc iharm} simulations by \cite{GammieIHARM2003,2021JOSS....6.3336P} from the standard library of 3D time-dependent GRMHD simulations performed in \citet{PaperV,PaperVIII}. 

\section*{acknowledgement}
It is a great pleasure to thank Peter Galison and Michael Johnson for very fruitful conversations. We are very grateful to the referee for very constructive comments. Razieh Emami acknowledges the support by the Institute for Theory and Computation at the Center for Astrophysics as well as grant numbers 21-atp21-0077, NSF AST-1816420 and HST-GO-16173.001-A. We thank the supercomputer facility at Harvard University where most of the simulation work was done. GNW gratefully acknowledges support from the Taplin Fellowship. ARR and KC acknowledge support by the National Science Foundation under Grant No. OISE 1743747. RJA acknowledges the Future Faculty Leaders Postdoctoral Fellowship. This research was made possible through the support of grants from the Gordon and Betty Moore Foundation and the John Templeton Foundation. 
Dominic Chang acknowledges the support of the Black Hole Initiative at Harvard University, which is funded by grants from the John Templeton  Foundation and the Gordon and Betty Moore Foundation to Harvard University. AC was supported by Hubble Fellowship grant HST-HF2-51431.001-A awarded by the Space Telescope Science Institute, which is operated by the Association of Universities for Research in Astronomy, Inc., for NASA, under contract NAS5-26555. AC also gratefully acknowledges support from the Princeton Gravity Initiative. This work was also supported by the National Science Foundation grants AST 1935980 and AST 2034306 and the Gordon and Betty Moore Foundation (GBMF-5278). The opinions expressed in this publication are those of the author(s) and do not necessarily reflect the views of the Moore or Templeton Foundations. Freek Roelofs was supported by NSF grants AST-1935980 and AST-2034306. I. Mart\'i-Vidal acknowledges support from the GenT program of Generalitat Valenciana (CIDEGENT/2018/021), MICINN Research Project PID2019-108995GB-C22 and the Astrophysics and High Energy Physics Programme by MCIN, with funding from European Union NextGenerationEU (PRTR-C17I1) and the Generalitat Valenciana through grant ASFAE/2022/018..

\begin{figure*}[th!]
\center
\includegraphics[width=0.99\textwidth]{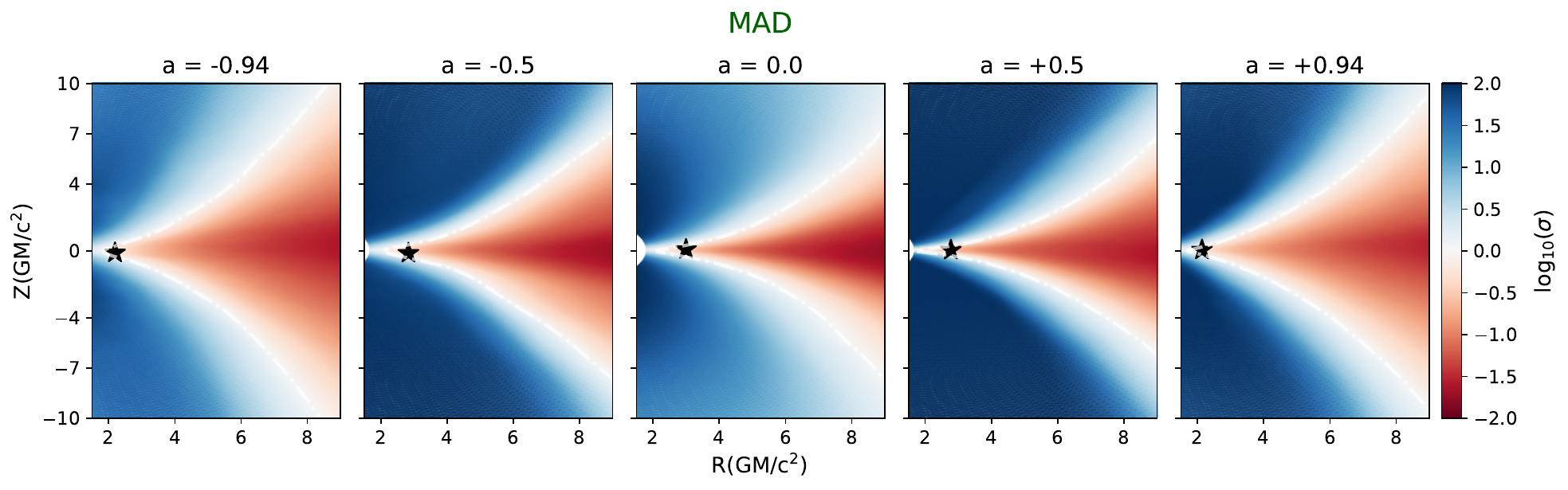}
\includegraphics[width=0.99\textwidth]{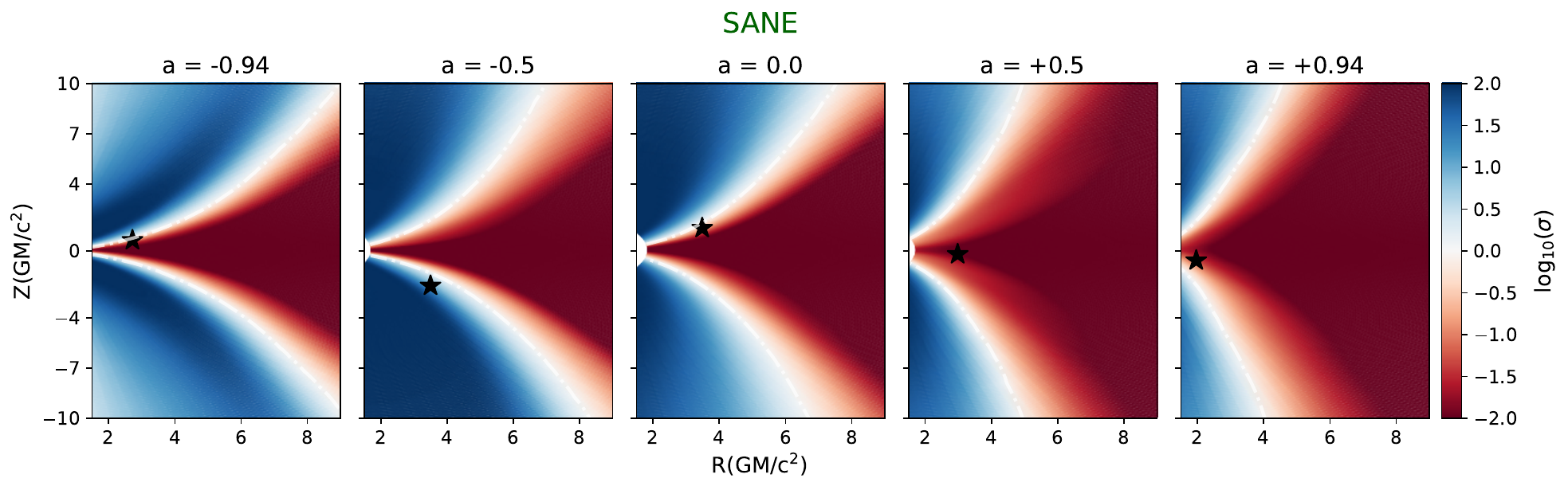}
\caption{The color-plot of $\log_{10}{(\sigma)}$ in the source $R$ vs $Z$ plane for MAD (top) and SANE (bottom) simulations.}
 \label{Sigma-R-Z-plot}
\end{figure*}

\textit{Software:} matplotlib \citep{2007CSE.....9...90H}, numpy \citep{2011CSE....13b..22V}, scipy \citep{2007CSE.....9c..10O}, seaborn \citep{2020zndo...3629446W}, pandas \citep{2021zndo...5203279R}, h5py \citep{2016arXiv160804904D}.

\appendix

\section{$\sigma$ contours in the $R$ vs $Z$ plane} \label{sigma-contours}
As already discussed in the main text, we made the BH images using $\sigma_{\mathrm{cut}}$ =1. This means that all the regions with a $\sigma \geq 1$ should be considered with a bit of caution. In 
Figure \ref{Sigma-R-Z-plot}, we present the color-plot of $\log_{10}{(\sigma)}$ in the source $R$ vs $Z$ plane using the time-averaged GRMHD simulations of MAD (top) and SANE (bottom) rows, respectively. In each panel the BH spin is increased in $a=(-0.94, -0.5, 0.0, +0.5, +0.94)$. Overlaid in each panel, we present the emission location with black star. From the plots, it is inferred that our emission location is fairly consistent with the $\sigma_{\mathrm{cut}}$ = 1.

\section{Averaged images from Faraday Thick model} \label{Image-Faraday-Thick}
\begin{figure*}[th!]
\center
\includegraphics[width=0.99\textwidth]{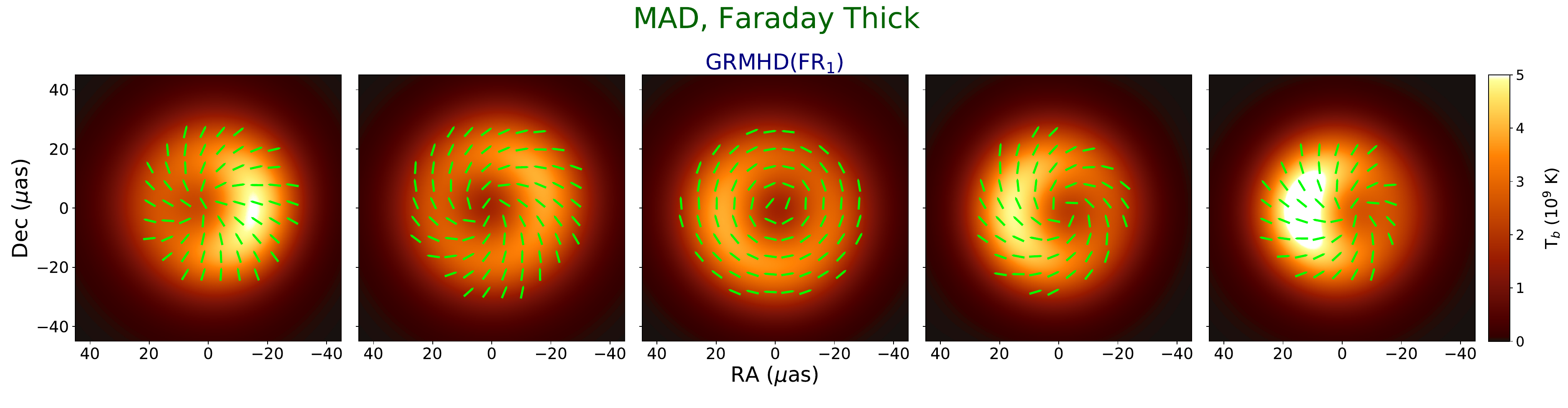}
\includegraphics[width=0.99\textwidth]{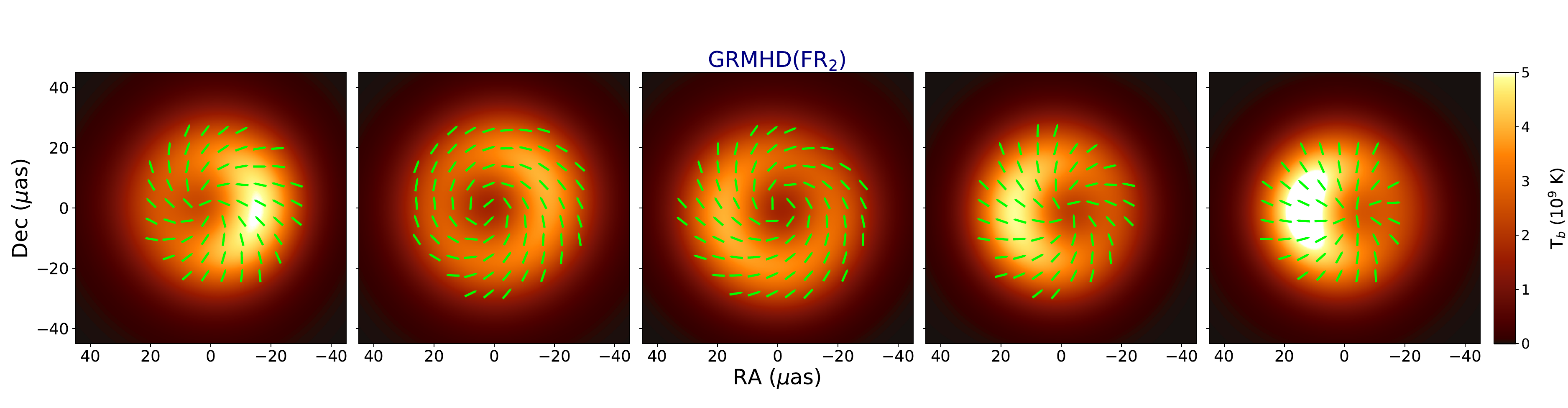}
\includegraphics[width=0.99\textwidth]{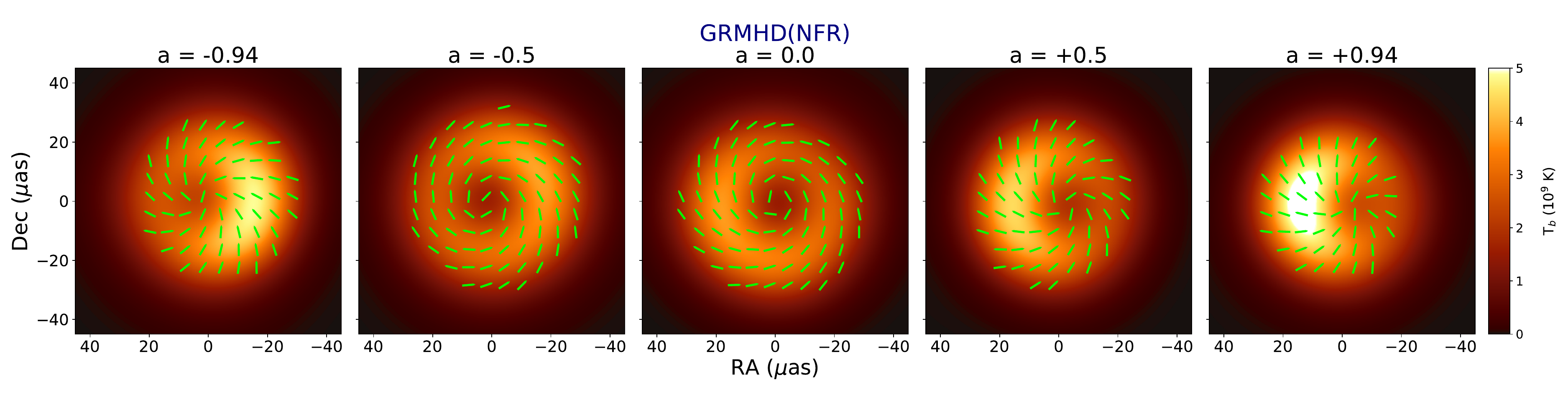}
\includegraphics[width=0.99\textwidth]{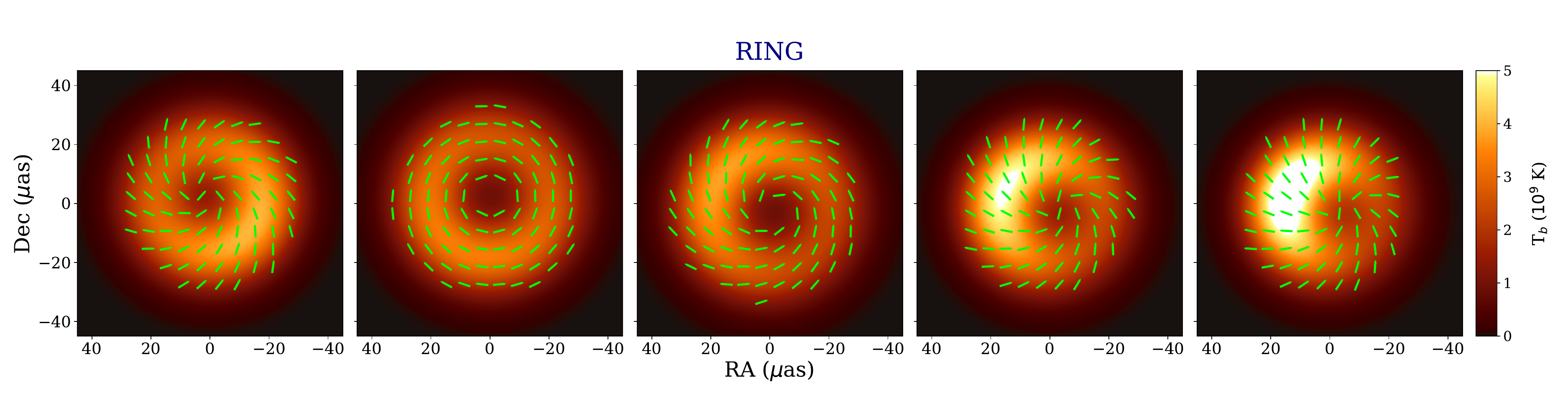}
\caption{The averaged image of intensity from MAD simulation which are Faraday Thick with $R_{\mathrm{low}} = 10$ and $R_{\mathrm{high}} = 160$ with different BH spins represented in consecutive columns, $a=(-0.94, -0.5, 0.0, +0.5, +0.94)$. The green tick lines refer to the EVPAs.}
 \label{mad-averaged-FT}
\end{figure*}

\begin{figure*}[th!]
\center
\includegraphics[width=0.99\textwidth]{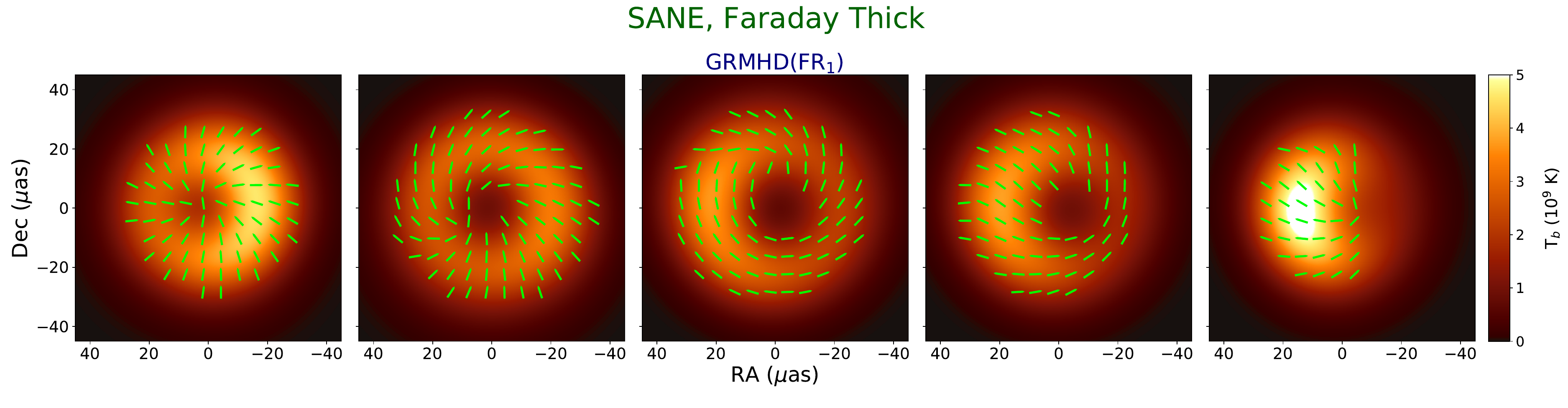}
\includegraphics[width=0.99\textwidth]{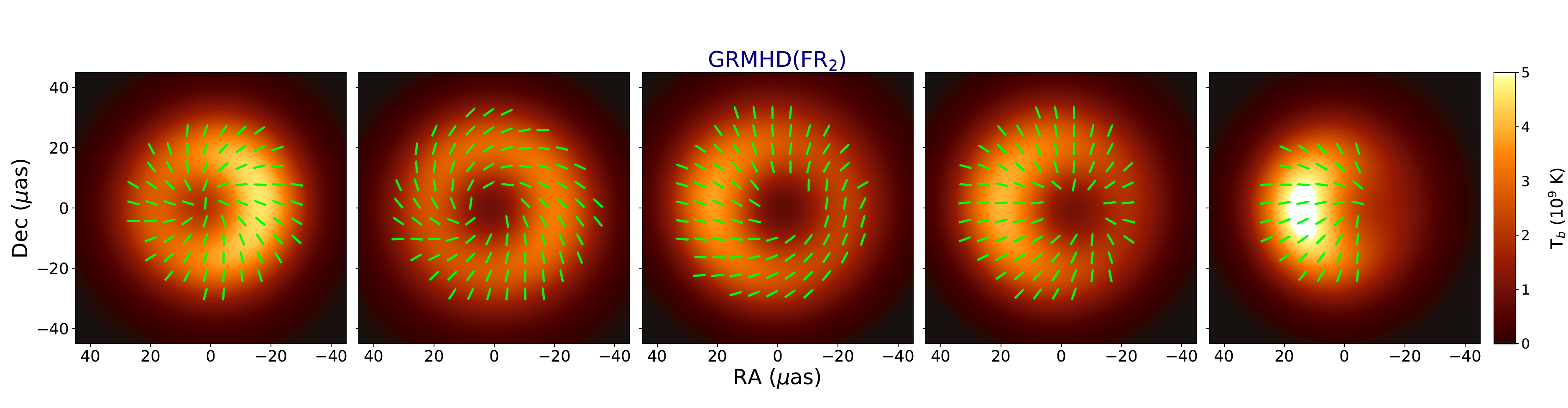}
\includegraphics[width=0.99\textwidth]{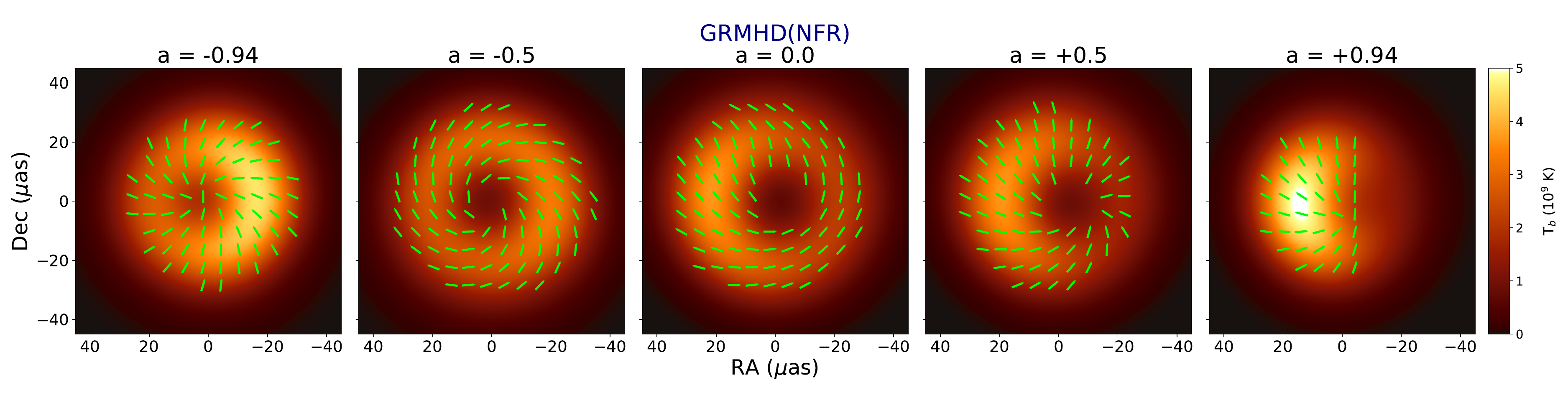}
\includegraphics[width=0.99\textwidth]{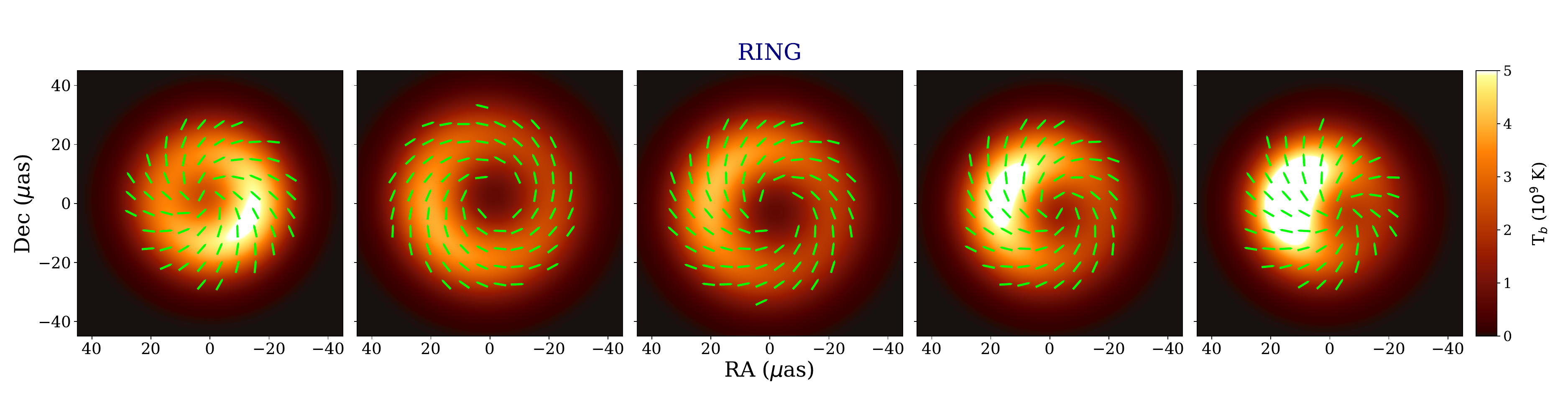}
\caption{The averaged image of intensity from the case of SANE simulation with Faraday Thick associated with $R_{\mathrm{high}} = 160$ and $R_{\mathrm{low}} = 10$ with different BH spins represented in consecutive columns, $a=(-0.94, -0.5, 0.0, +0.5, +0.94)$. The green tick lines refer to the EVPAs.}
 \label{Sane-averaged-FT}
\end{figure*}

Having presented the time averaged images of the Fiducial case in Figures \ref{Mad-averaged-fiducial} and \ref{Sane-averaged-fiducial}, here we focus on Faraday Thick models. Figures \ref{mad-averaged-FT} and \ref{Sane-averaged-FT} present the time averaged images for MAD and SANE models with different BH spins, respectively. 
Descending rows present $FR_1$, $FR_2$, NFR and the ring model. Comparing these images with the Fiducial case, it is evident that the Faraday rotation is more important here than there. Consequently, the level of the agreement with the ring model diminishes in the Faraday Thick case than the Fiducial model. Overall, MAD simulations provide a better agreement to the geometrical ring model than the SANE ones. In both cases, the more radially oriented EVPAs at higher absolute BH spin, gets more azimuthally oriented in low/zero spin models.

\section{Impact of the de-boosting and de-lensing}

To isolate different contributions in the ring model, in the following, we assess versions of the ring model in which either the plasma boost factor or the gravitational  lensing term are turned off. The Fiducial and the Faraday Thick cases are considered separately.  In order to diagnose the effects of Doppler beaming on the observed EVPA distribution, we utilize the ray traced toy model while setting the fluid velocity to zero in Boyer-Lindquist coordinates. On the other hand, to study the contribution of the general-relativistic lensing, we scale down the BH mass in the ray-tracing code while scaling the radius of emission in gravitational coordinates up, maintaining the same physical size of the emission and the same approximate angular size of the ring. Furthermore, we preserve the velocity structure of the simulation, effectively performing ray-tracing in a flat spacetime (de-lensed test).

\subsection{Fiducial case}
\begin{figure*}[th!]
\center
\includegraphics[width=0.99\textwidth]{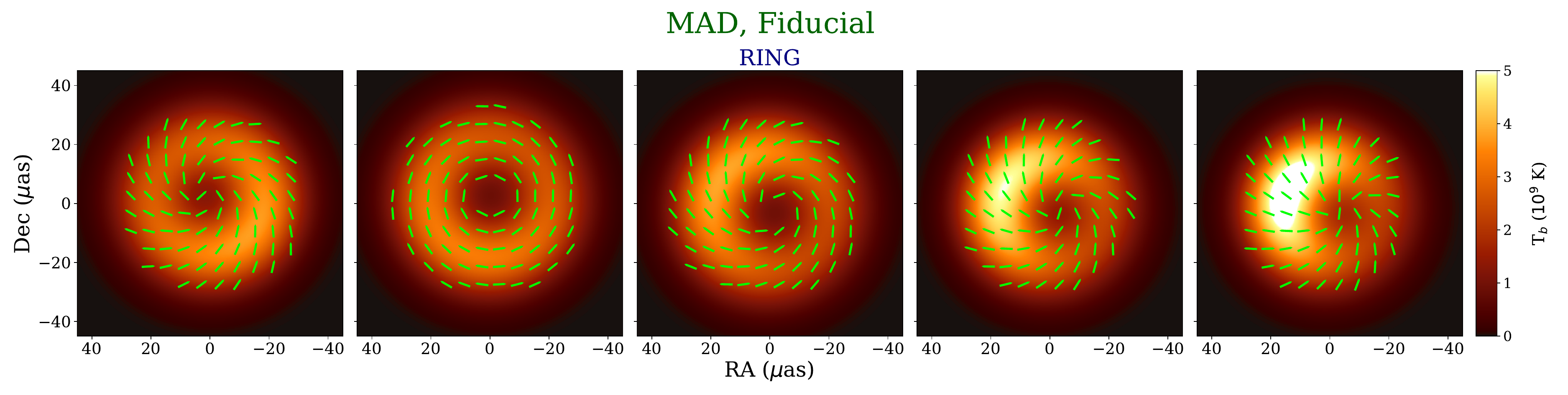}
\includegraphics[width=0.99\textwidth]{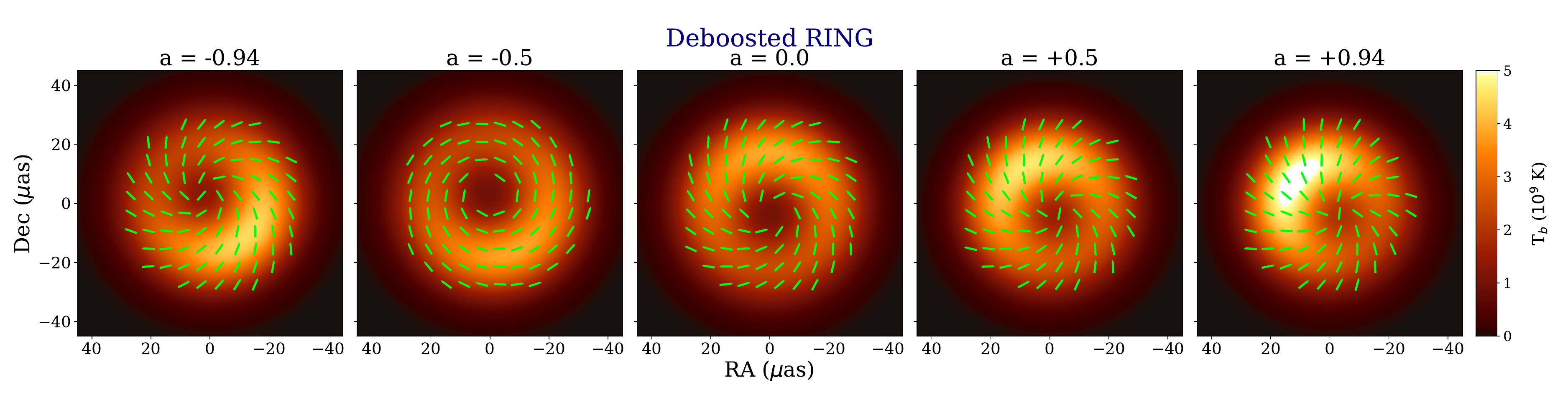}
\includegraphics[width=0.99\textwidth]{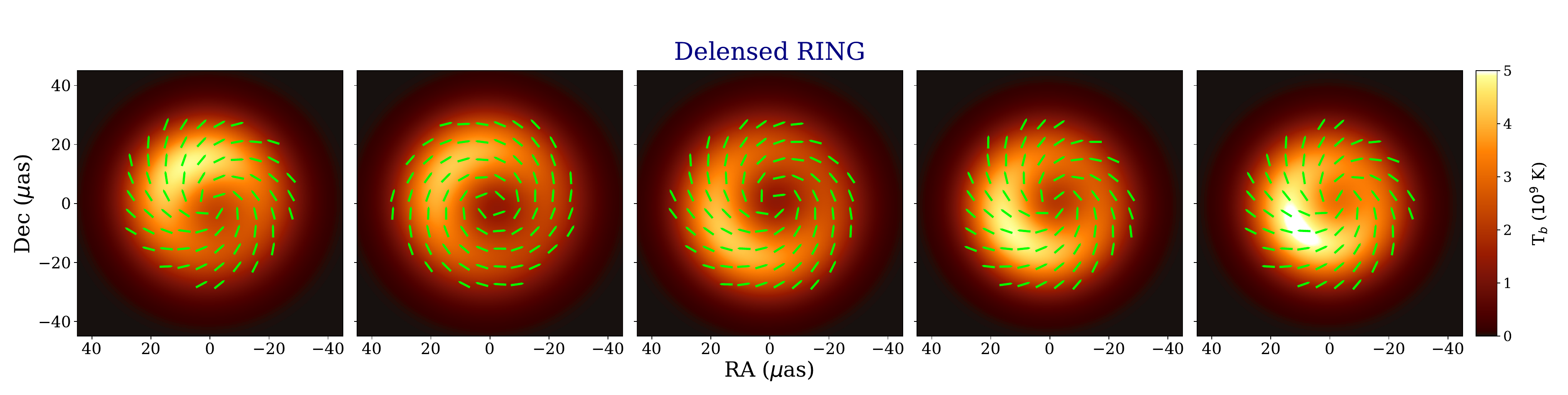}
\caption{The impact of turning off the plasma boost as well as the lensing in the Fiducial case for MAD simulations. From the top row to the bottom one, we present the full ring, deboosted and the delensed cases, respectively, with different spin values represented in consecutive columns.}
 \label{Mad-Deboost-Delensed}
\end{figure*}

\begin{figure*}[th!]
\center
\includegraphics[width=0.99\textwidth]{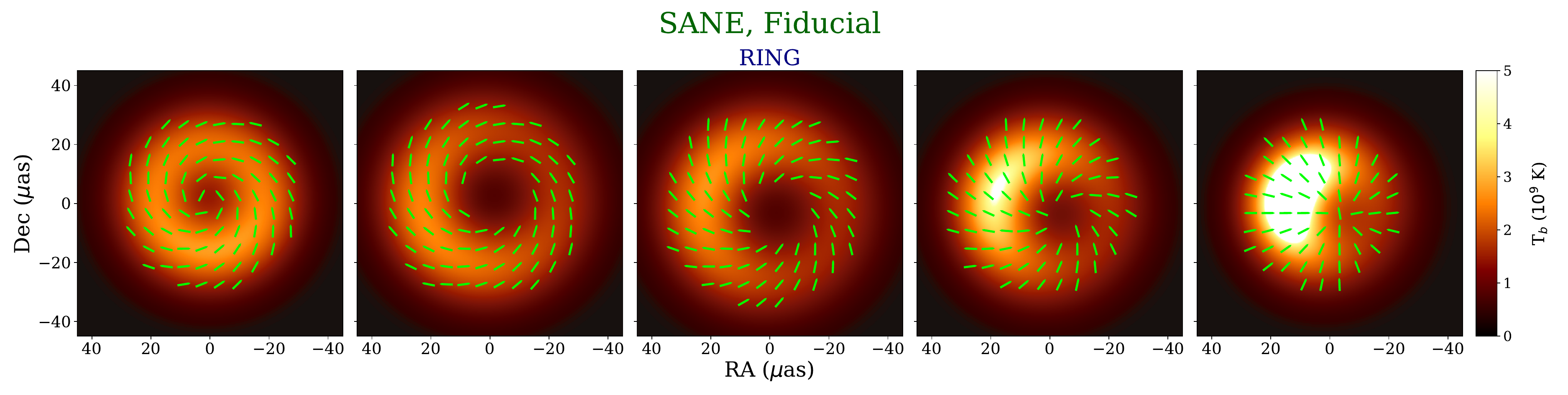}
\includegraphics[width=0.99\textwidth]{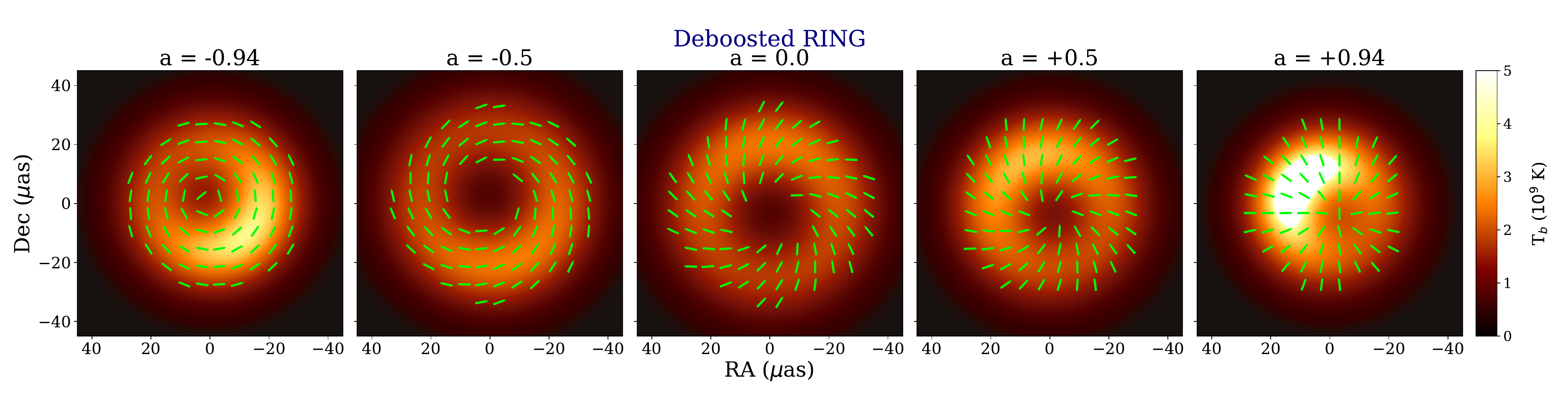}
\includegraphics[width=0.99\textwidth]{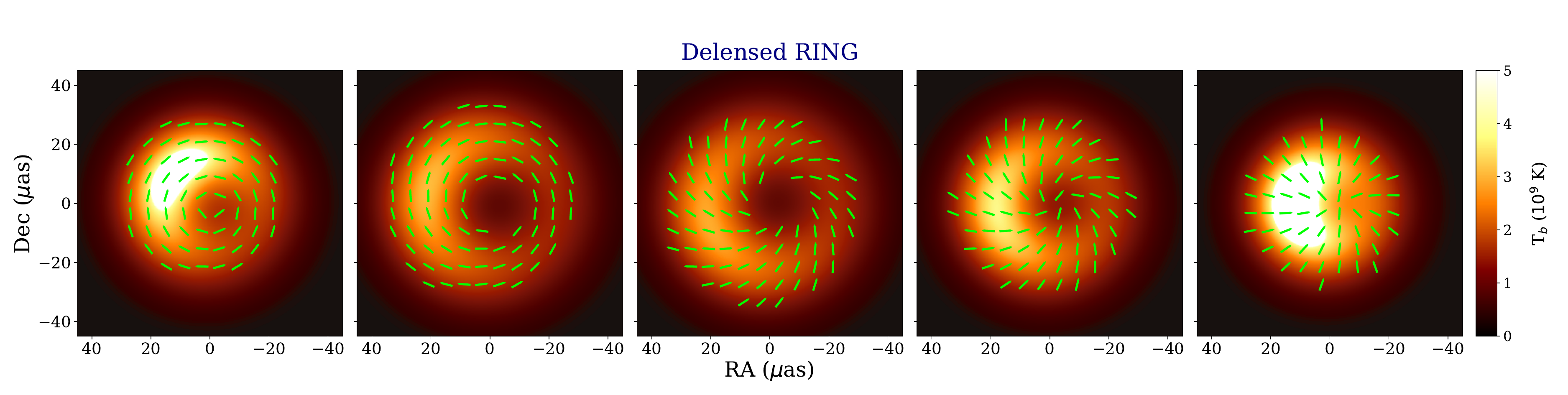}
\caption{The impact of turning off the plasma boost as well as the lensing in the Fiducial case for SANE simulations. From the top row to the bottom one, we present the full ring, deboosted and the delensed cases, respectively, with different spin values represented in consecutive columns.}
 \label{SANE-Deboost-Delensed}
\end{figure*}

Here we analyse the Fiducial simulations. Figures \ref{Mad-Deboost-Delensed} and \ref{SANE-Deboost-Delensed} compare the polarized images from the ring model (top row) with the deboosted ring (middle row) and the delensed ring (bottom row) for models of MAD and SANE in the Fiducial case, respectively. From the plots, it is inferred that:

In MAD simulations, the brightness distribution of the original ring is very similar to the deboosted ring model. This is however not the case for the delensed ring model in which we see substantial differences between the brightness distribution of the original ring compared with the delensed ring model. The ticks of the EVPAs, on the other hand, seems more similar between different cases in MAD simulations. The details of the comparison between the ticks of EVPAs ca be find in Figure \ref{Ring-Deboosting-Delensing}.

In SANE simulations, the differences between the brightness distribution of the ring model (top row) and the 
deboosted ring (middle row) is more noticeable with more differences in the retrograde and zero spin than the prograde cases. The delensed ring model, on the other hand, shows more similarity to the ring model (top row) with an only exception for the case with $a = -0.94$. There are some levels of changes between the ticks of the EVPAs in all of these cases as well. We have quantified such differences in Figure \ref{Ring-Deboosting-Delensing}.

\subsection{Faraday Thick case}
\begin{figure*}[th!]
\center
\includegraphics[width=0.99\textwidth]{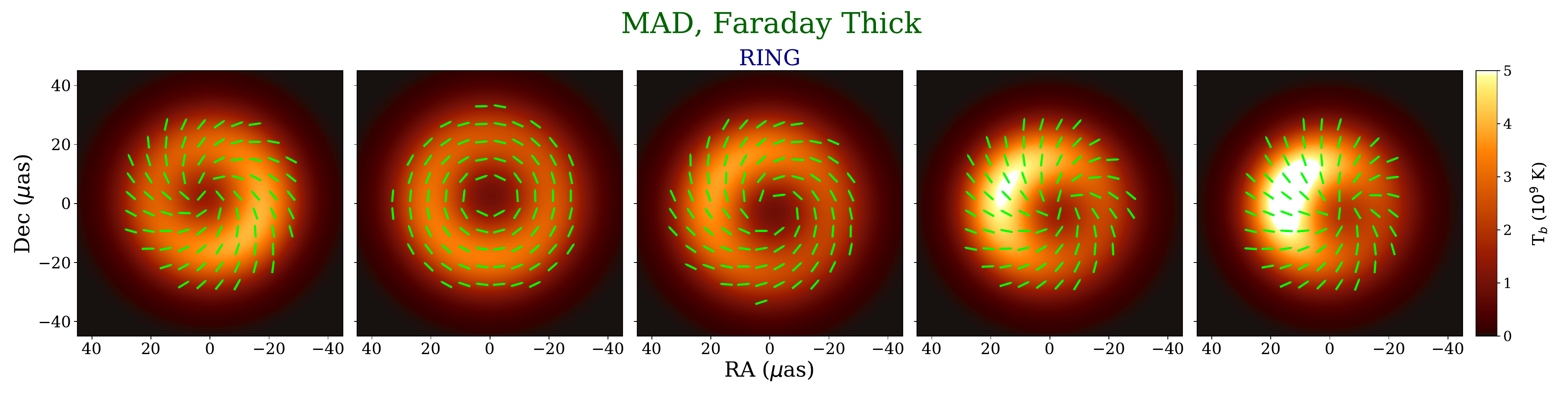}
\includegraphics[width=0.99\textwidth]{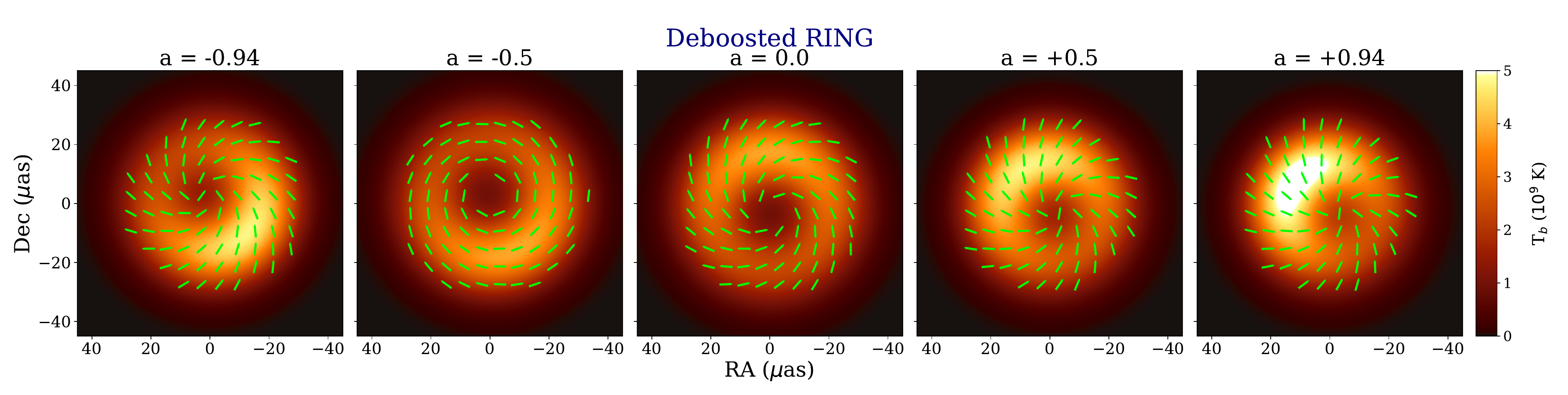}
\includegraphics[width=0.99\textwidth]{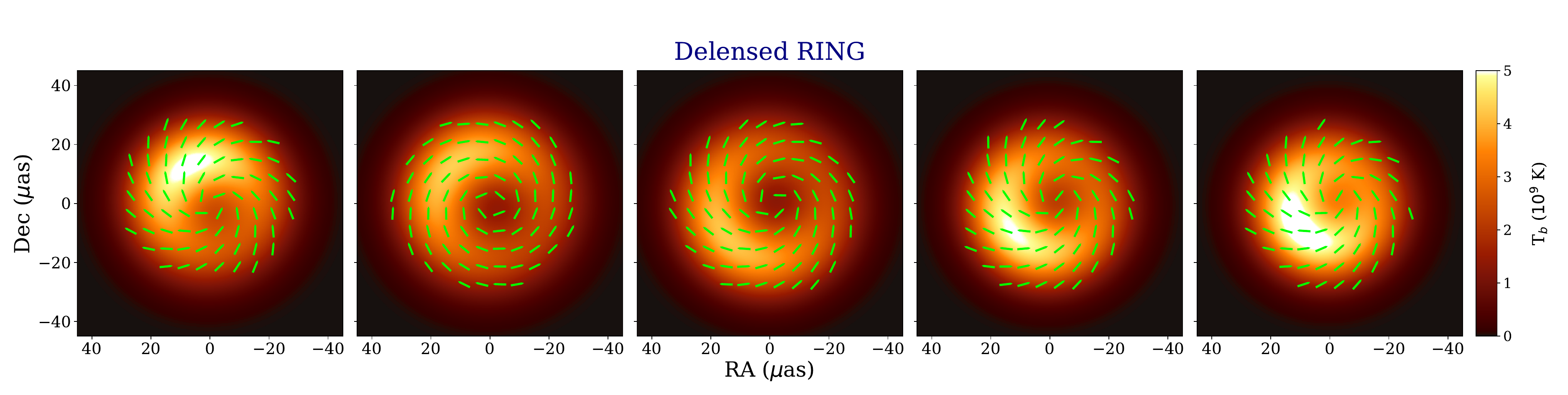}
\caption{The impact of turning off the plasma boost as well as the lensing in the Faraday Thick case for MAD simulations. From the top row to the bottom one, we present the full ring, deboosted and the delensed cases, respectively, with different spin values represented in consecutive columns.}
 \label{Mad-Deboost-Delensed160}
\end{figure*}

\begin{figure*}[th!]
\center
\includegraphics[width=0.99\textwidth]{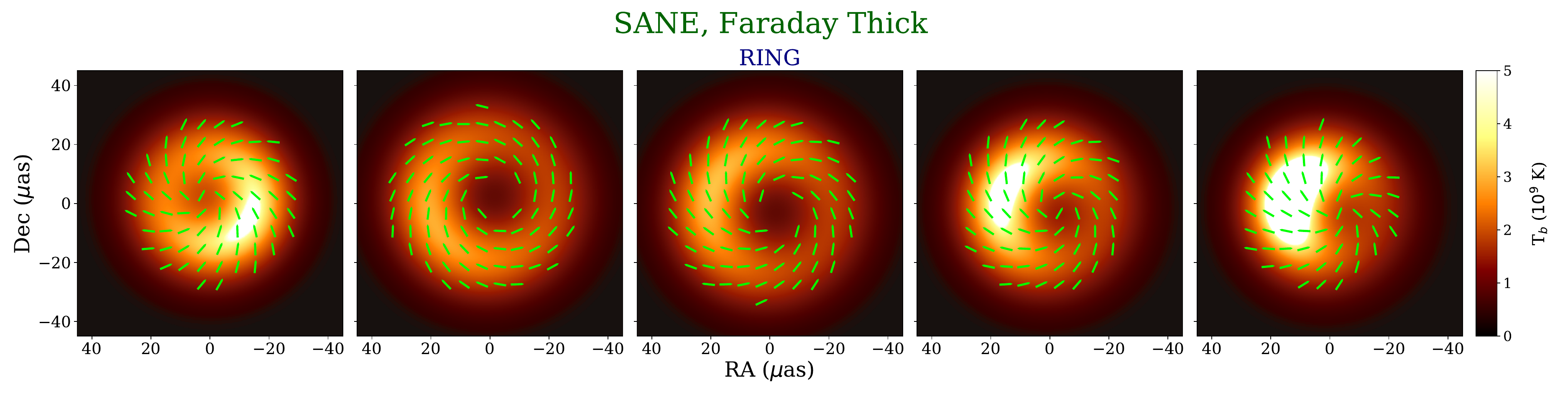}
\includegraphics[width=0.99\textwidth]{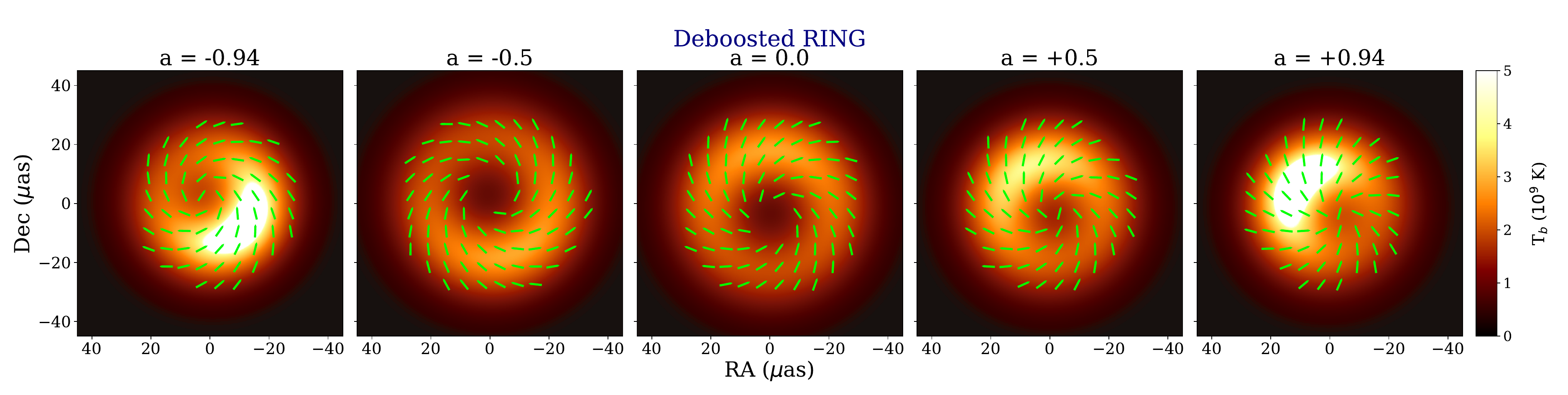}
\includegraphics[width=0.99\textwidth]{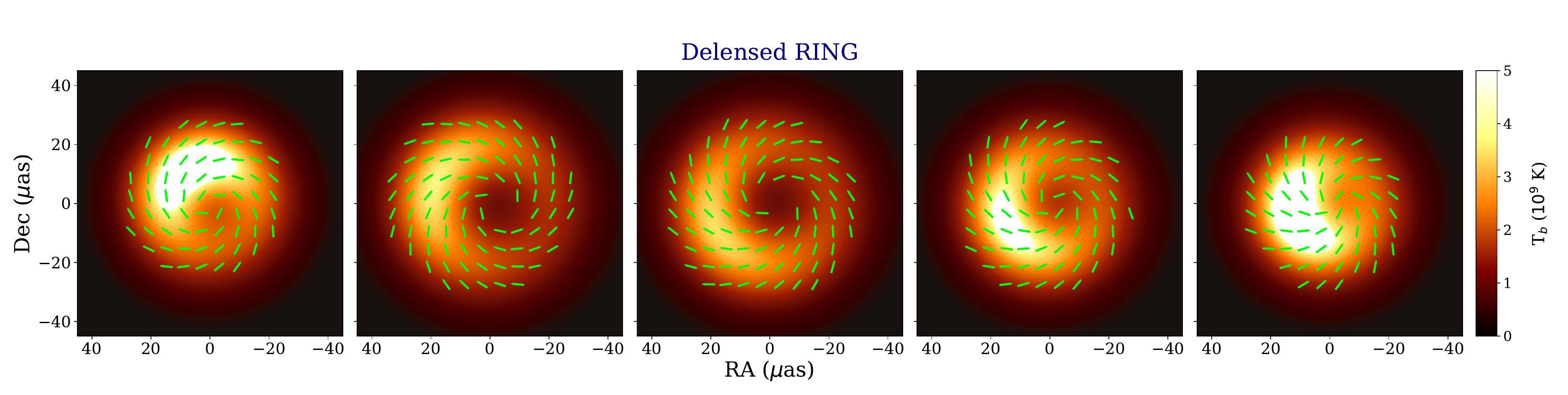}
\caption{The impact of turning off the plasma boost as well as the lensing in the Faraday Thick case for SANE simulation. From the top row to the bottom one, we present the full ring, deboosted and the delensed cases, respectively, with different spin values represented in consecutive columns.}
 \label{SANE-Deboost-Delensed160}
\end{figure*}

Next, we study the Faraday Thick simulations. Figures \ref{Mad-Deboost-Delensed160} and \ref{SANE-Deboost-Delensed160} compare the polarized images from the ring model (top row) with the deboosted ring (middle row) and the delensed one (bottom row) for models of MAD and SANE in the Faraday Thick case, respectively. From the plots, it is inferred that:

In MAD simulations, both the brightness distribution of the deboosted ring model and the ticks of the EVPAs follows closely the behavior of the original ring model, which is quite similar to the above Fiducial case, as seen in Figure \ref{Mad-Deboost-Delensed}. The structure however is rather different in the delensed ring model. A more quantitative comparison between the ticks of the EVPAs can be found in the bottom row in Figure \ref{Ring-Deboosting-Delensing}.

In SANE simulations, the distribution of the brightness in the deboosted case shows a bit less agreement with the original ring model than in MAD simulations. The ticks of the EVPAs also show slightly less agreement for retrograde spins. The brightness distribution of the delensed models show substantial differences compared with the original ring model. This behavior is quite similar to the Fiducial case, as shown in Figure \ref{SANE-Deboost-Delensed}. The details of the quantitative comparison between these models can be found in Figure \ref{Ring-Deboosting-Delensing}.

\section{Effect of the Magnetic Field Polarity}
\label{sec:flip_b}

Throughout this work, we have considered two orientations of the magnetic field polarity, either aligned or anti-aligned with disk angular momentum on large scales.  EHT science thus far has focused on models with the magnetic field aligned with the disk angular momentum.  However, since the equations of GRMHD are invariant to a flip in the magnetic field direction, the same GRMHD snapshots can be ray-traced using either magnetic field polarity.  This has negligible impact on total intensity, but can alter both the linear and circular polarization structure.  Depending on the model, flipping the magnetic field direction can result in a complicated change in the circular polarization structure, since intrinsic emission and Faraday rotation change sign, but Faraday conversion does not \citep[][]{Ricarte+2021}.  In linear polarization, the direction of the field causes a systematic rotation of polarization ticks due to Faraday rotation, which directly impacts $\angle \beta_2$.

\begin{figure*}
\center
\includegraphics[width=0.98\textwidth]{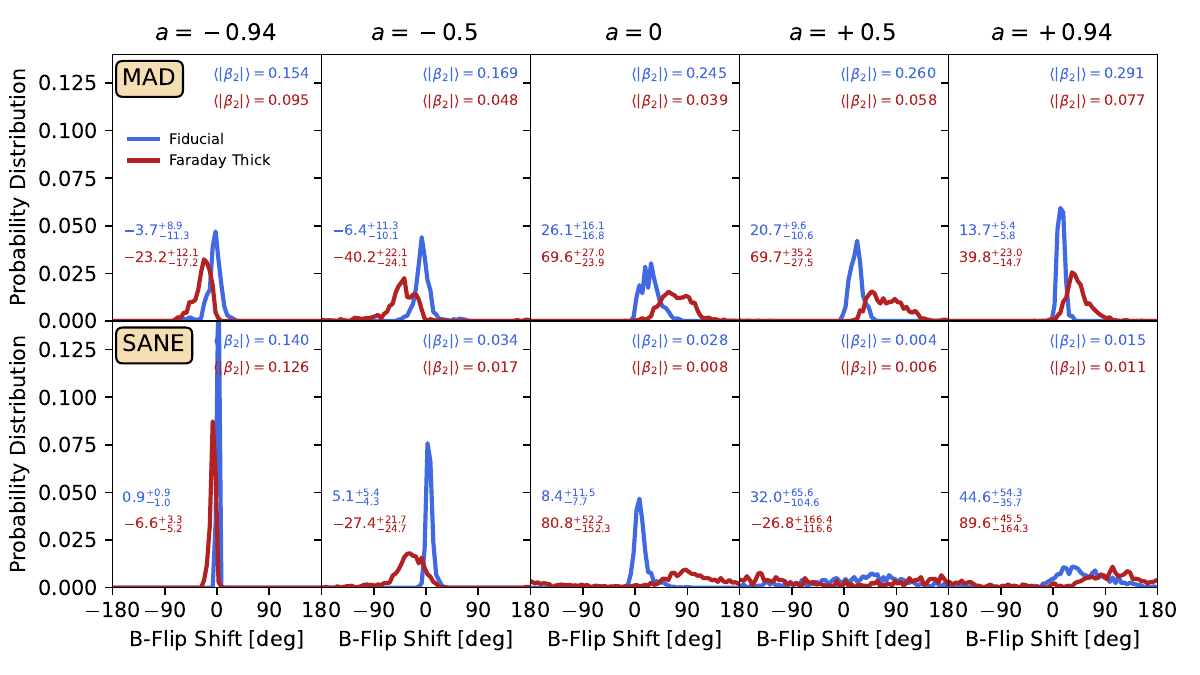}
\caption{Differences in $\angle \beta_2$ between our anti-aligned and aligned magnetic field models ($\angle \beta_2(\mathrm{FR}_2)-\angle \beta_2(\mathrm{FR}_1)$).  The effect of flipping the magnetic field is stronger for more Faraday Thick models, as expected.  We find negative shifts in $\angle \beta_2$ for retrograde models and positive shifts for the others because these two sets of models are initialized with opposite magnetic field polarity by construction.}
 \label{fig:flip_b}
\end{figure*}

In \autoref{fig:flip_b}, we plot the distribution of changes in $\angle\beta_2$ upon flipping the field ($\angle \beta_2(\mathrm{FR}_2)-\angle \beta_2(\mathrm{FR}_1)$) on a snapshot-by-snapshot basis.  As expected, the effect is stronger for models with more Faraday rotation.  This flip induces positive shifts in $\angle \beta_2$ for prograde and spin 0 models and negative shifts in retrograde models because their poloidal magnetic fields are oriented in opposite directions by construction.  We expect that the observable quantities sensitive to the magnetic field polarity are $\angle \beta_2$, circular polarization, and rotation measure.

\section{Distributions of $|\beta_2|$}
\label{sec:amp_beta2}

\begin{figure*}[th!]
\center
\includegraphics[width=0.98\textwidth]{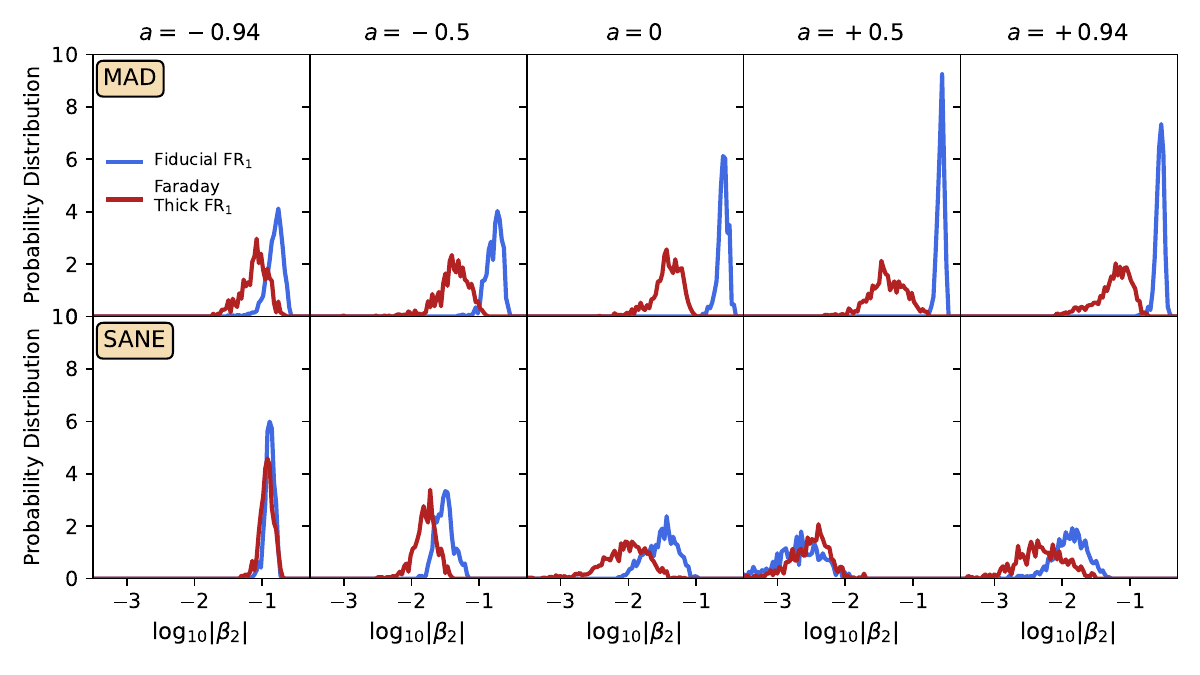}
\caption{Distributions of $|\beta_2|$ for our models with aligned magnetic fields.  Faraday rotation drives the most important trends: both SANE models and models with larger values of $R_\mathrm{high}$ and $R_\mathrm{low}$ have increased Faraday depths, which scrambles $|\beta_2|$.}
 \label{fig:amp_beta2}
\end{figure*}

Throughout this manuscript, we have focused on $\angle \beta_2$ rather than $|\beta_2|$.  Our analytic ring model can make no meaningful prediction for $|\beta_2|$ because it does not include radial evolution nor does it account for Faraday rotation. We plot the distributions of $|\beta_2|$ for our Fiducial and Faraday Thick (aligned magnetic field) models in \autoref{fig:amp_beta2}.  As discussed in previous works \citep{Palumbo_2020,PaperVIII}, $|\beta_2|$ is larger for MAD models than for SANE models.  In addition, as we had expected when making these model sets, our Faraday Thick models have lower $|\beta_2|$ than the Fiducial models.  As discussed in \citet{PaperVIII}, the $a=-0.94$ SANE models have uncharacteristically large $|\beta_2|$ because much of the polarization signal comes from the forward jet, which avoids the large Faraday rotation depth in the mid-plane.

\newpage

\bibliography{main}

\end{document}